%% file: Hyperbolic09.tex
\documentclass[notitlepage,superscriptaddress,onecolumn,english,prl,aps]{revtex4-1}
\usepackage[T1]{fontenc}
\usepackage[latin9]{inputenc}
\usepackage{amsmath}
\usepackage{amssymb}
\usepackage{graphicx}
\usepackage{hyperref}
\usepackage{docmute}
\makeatletter

\usepackage{mathrsfs}

\makeatother

\usepackage{babel}
\begin{document}

\title{NV-Metamaterial: Tunable Quantum Hyperbolic Metamaterial Using Nitrogen-Vacancy Centers
in Diamond}

\author{Qing Ai}

\affiliation{CEMS, RIKEN, Wako-shi, Saitama 351-0198, Japan}

\affiliation{Department of Physics, Applied Optics Beijing Area Major Laboratory,
Beijing Normal University, Beijing 100875, China}

\author{Peng-Bo Li}

\affiliation{CEMS, RIKEN, Wako-shi, Saitama 351-0198, Japan}

\affiliation{Department of Applied Physics, Xi\textquoteright an Jiaotong University,
Xi\textquoteright an 710049, China}

\author{Wei Qin}

\affiliation{CEMS, RIKEN, Wako-shi, Saitama 351-0198, Japan}

\affiliation{Quantum Physics and Quantum Information Division, Beijing Computational
Science Research Center, Beijing 100193, China}

\author{C. P. Sun}

\affiliation{Beijing Computational Science Research Center \&
Graduate School of Chinese Academy of Engineering Physics, Beijing 100084, China}

\author{Franco Nori}

\affiliation{CEMS, RIKEN, Wako-shi, Saitama 351-0198, Japan}

\affiliation{Department of Physics, The University of Michigan, Ann Arbor, Michigan
48109-1040, USA}

\begin{abstract}
We show that nitrogen-vacancy (NV) centers in diamond can produce
a novel quantum hyperbolic metamaterial. We demonstrate that a hyperbolic
dispersion relation in diamond with NV centers can be engineered and
dynamically tuned by applying a magnetic field. This quantum hyperbolic
metamaterial with a tunable window for the negative refraction allows
for the construction of a superlens beyond the diffraction limit.
In addition to subwavelength imaging, this NV-metamaterial can be used in spontaneous emission enhancement, heat transport and acoustics, analogue cosmology, and lifetime engineering.
Therefore, our proposal interlinks the two hotspot fields, i.e., NV centers and metamaterials.
\end{abstract}
\maketitle

\textit{Metamaterials}.\textendash{}\textendash{}Metamaterials with negative refraction have attracted broad interest \cite{Bliokh2008,Veselago1968,Pendry2000,Smith2000}.
Metamaterials can be used, e.g., for electromagnetic cloaking,
perfect lens beyond diffraction limit \cite{Pendry2000}, fingerprint
identification in forensic science \cite{Shen2016},
simulating condensate matter phenomena \cite{Bliokh2013} and
reversed Doppler effect \cite{Kats2007}. In order to realize
negative refraction, sophisticated composite architectures \cite{Smith2000,Yao2008}
and topologies \cite{Fang2016,Rakhmanov2010,Chang2010,Shen2014} are fabricated
to achieve simultaneously negative permittivity and permeability.
However, hyperbolic (or indefinite) metamaterials were proposed
\cite{Poddubny2013,Jahani2016,Smith2003,Belov2003} to overcome the
difficulty of inducing a magnetic transition at the same frequency
as the electric response. The magnetic response of double-negative metamaterials is so
weak that it effectively shortens the frequency window of the negative
refraction \cite{Fang2016}. In addition to subwavelength imaging \cite{Liu2007,Ishii2013} and focusing \cite{Ishii2013}, hyperbolic metamaterials have been used to
realize spontaneous emission enhancement \cite{Jacob2012},
applied in heat transport \cite{Biehs2012} and acoustics \cite{Li2009},
analogue cosmology \cite{Smolyaninov2010}, and lifetime engineering \cite{Krishnamoorthy2012,Yang2012}.

\textit{NV centers}.\textendash{}\textendash{}On the other hand,
quantum devices based on nitrogen-vacancy (NV) centers
in diamond are under intense investigation \cite{Schirhagl2014,Wu2016}
as they manifest some novel properties
and can be explored for many interesting applications \cite{Doherty2013,Chen2017}.
For example, NV centers in diamond have been proposed to realize a laser \cite{Jeske2017}
and maser \cite{Jin2015} at room temperature. Highly-sensitive solid-state
gyroscopes \cite{Ledbetter2012} based on ensembles of NV centers in diamond can be realized
by dynamical decoupling, to suppress the dipolar relaxation.
Shortcuts to adiabaticity have been successfully performed
in NV centers of diamond to initialize and transfer coherent superpositions
\cite{Zhou2017,Song2016}.
The high sensitivity to external signals makes single NV centers
promising for quantum sensing of various physical parameters, such as
electric field \cite{Dolde2011,Dolde2014}, magnetic field \cite{Maze2008,Balasubramanian2008,Li2017}, single electron and nuclear spin \cite{Zhao2011,Grinolds2013,Cooper2014,Shi2015,DeVience2015,Boss2016,Liu2017,Wang2017}, and temperature \cite{Kucsko2013,Toyli2013,Neumann2013}.
Numerous hybrid quantum devices, composed of NV
centers and other quantum systems, e.g. superconducting circuits and
carbon nanotubes, have been proposed to realize demanding tasks \cite{Xiang2013-1,Xiang2013-2,Lv2013,Li2016,Zagoskin2007}.

\textit{NV-metamaterials}.\textendash{}\textendash{}Inspired by the rapid progress in both fields,
here we propose to realize a hyperbolic metamaterial using NV centers
in diamond. We consider an electric hyperbolic metamaterial,
in which two principal components of its electric permittivity possess
different signs. When an optical electromagnetic field induces the
transition $^{3}A_{2}\rightleftharpoons{}^{3}E$, the NV centers
in diamond will negatively respond to the electric field in one direction.
This process effectively modifies the relative permittivity of the
diamond with NV centers and thus one principal component
has a different sign. When a transverse magnetic (TH) mode is incident
on this diamond with the principal axis of the negative component perpendicular
to the interface, the transmitted light will be negatively refracted,
as both the incident and transmitted light lie at the same side of
the normal to the interface.
Note that it is difficult to fabricate classical metamaterials
working in the optical-frequency domain,
because the sizes of the elements therein are sub-micron.
However, the NV centers in diamond can be easily fabricated in several ways \cite{Doherty2013},
e.g., as an in-grown product of the
chemical vapour deposition diamond synthesis process,
as a product of radiation damage and annealing,
as well as ion implantation and annealing in bulk and nanocrystalline diamond.
The NV-metamaterials proposed here solve this problem.

\begin{figure}

\includegraphics[bb=110 300 500 490,width=8.5cm]{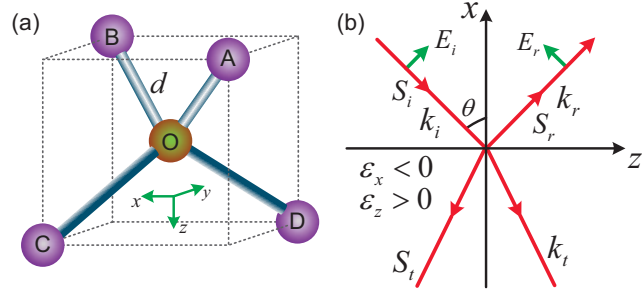}
\caption{(color online) (a) Four possible orientations of NV centers in diamond \cite{Doherty2013,Zou2014}:
$\vec{r}_\textrm{OA}=(-1,-1,-1)/\sqrt{3}$, $\vec{r}_\textrm{OB}=(1,1,-1)/\sqrt{3}$, $\vec{r}_\textrm{OC}=(1,-1,1)/\sqrt{3}$,
$\vec{r}_\textrm{OD}=(-1,1,1)/\sqrt{3}$. $d=154$~pm is the length of carbon
bond. The angle between any pair of the above four orientations is identically
$\alpha=109\protect\textdegree28'$. (b) Negative refraction for
hyperbolic dispersion with $\epsilon_{x}<0$ and $\epsilon_{z}>0$.
The TH mode is incident on the $yz$ interface with electric field $\vec{E}_{i}$,
wavevector $\vec{k}_{i}$, and Poynting vector $\vec{S}_{i}$. The angle
between the normal ($x$-axis) and the incident field is $\theta$. It
is reflected with electric field $\vec{E}_{r}$, wavevector $\vec{k}_{r}$,
and Poynting vector $\vec{S}_{r}$. The Poynting and wavevector
of the transmitted wave are, respectively, $\vec{S}_{t}$ and $\vec{k}_{t}$.\label{fig:scheme}}
\end{figure}

\textit{Model}.\textendash{}\textendash{}As schematically illustrated in Fig.~\ref{fig:scheme}(a),
an NV center is composed of a vacancy, e.g. site O, and a substitutional
nitrogen atom at one of its four possible neighboring sites, e.g.
site A, B, C and D. The electronic ground state is a spin-triplet
state with Hamiltonian \cite{Doherty2013,SuppMat}
\begin{eqnarray}
H_{\mathrm{gs}} & \!\!=\!\! & D_{\mathrm{gs}}S_{z}^{2}+\mu_{B}g_{\mathrm{gs}}^{\parallel}B_{z}S_{z}+\mu_{B}g_{\mathrm{gs}}^{\perp}(B_{x}S_{x}+B_{y}S_{y}),
\end{eqnarray}
where $D_{\mathrm{gs}}=2.88$~GHz is the zero-field splitting of the
electronic ground state, $\mu_{B}$ is the Bohr magneton, $g_{\mathrm{gs}}^{\parallel}\simeq g_{\mathrm{gs}}^{\perp}=g_{\mathrm{gs}}\simeq2$
are respectively the components of the $g$-factor of the electronic ground
state, $\vec{B}$ is the magnetic field, and $S_{\alpha}$ ($\alpha=x,y,z$)
are the spin-1 operators for the electron spin.

At room temperature, when there is no electric and strain fields,
the Hamiltonian of the electronic excited state is simplified as \cite{Doherty2013,SuppMat}
\begin{eqnarray}
H_{\mathrm{es}} & = & D_{\mathrm{es}}^{\parallel}S_{z}^{2}+\mu_{B}g_{\mathrm{es}}^{\mathrm{RT}}\vec{B}\cdot\vec{S}+\xi(S_{y}^{2}-S_{x}^{2}),
\end{eqnarray}
where $D_{\mathrm{es}}^{\parallel}=1.42$~GHz is the zero-field splitting
of the electronic excited state, $g_{\mathrm{es}}^{\mathrm{RT}}\simeq2.01$
is the $g$-factor of the electronic spin of the excited state at
room temperature, $\xi=70$~MHz is the strain-related coupling.

As illustrated in Fig.~\ref{fig:scheme}(a), there are four possible
orientations for the NV centers in diamond \cite{Schirhagl2014,Wu2016,Doherty2013,Chen2017,Zou2014,SuppMat}.
Since both Hamiltonians of the ground and excited states
are obviously dependent on the relative orientation
of the symmetry axis with respect to the magnetic field,
the energy spectra and thus the electromagnetic response of the NV
centers to the applied fields are different for the four possible
orientations.


\textit{Selection Rules of Optical Transitions}.\textendash{}\textendash{}According to Refs.~\cite{Doherty2011,Maze2011}, there are four outer electrons
distributed in the $a_{1}$, $e_{x}$ and $e_{y}$ levels, i.e. $a_{1}^{2}e^{2}$.
On account of the spin degree of freedom, the electronic ground states
are the triplet states labeled as \cite{SuppMat} $\vert\Phi_{A_{2};1,0}^{c}\rangle$,
$\vert\Phi_{A_{2};1,1}^{c}\rangle$, $\vert\Phi_{A_{2};1,-1}^{c}\rangle$,
where the superscript $c$ means configuration, the subscripts are
ordered as $j,k;S,m_{s}$ with $j$ being irreducible representation,
$k$ being row of irreducible representation, $S$ being total spin and $m_{s}$ being spin
projection along the symmetry axis of the NV center. The six first-excited
states, i.e. $a_{1}e^{3}$, are \cite{SuppMat} $\vert\Phi_{E,x;1,0}^{c}\rangle$,
$\vert\Phi_{E,y;1,0}^{c}\rangle$, $\vert\Phi_{E,x;1,1}^{c}\rangle$,
$\vert\Phi_{E,y;1,1}^{c}\rangle$, $\vert\Phi_{E,x;1,-1}^{c}\rangle$,
$\vert\Phi_{E,y;1,-1}^{c}\rangle$, where $\vert\Phi_{E,x;S,m_{s}}^{c}\rangle$
and $\vert\Phi_{E,y;S,m_{s}}^{c}\rangle$ are degenerate under a magnetic field.
By comparing the ground and excited states,
there is one electron transiting from the $a_{1}$
orbital to the $e$ orbital. Without spin-orbit coupling,
due to conservation of spin and total angular momentum \cite{Togan2010},
the non-zero transition matrix elements of the position vector $\vec{r}=x\hat{e}_{x}+y\hat{e}_{y}+z\hat{e}_{z}$ are in the following
transitions
$\vert\Phi_{A_{2};S,m_{s}}^{c}\rangle\stackrel{\alpha^{\prime}}{\rightleftharpoons}\vert\Phi_{E,\alpha;S,m_{s}}^{c}\rangle$ \cite{Maze2011,SuppMat},
where $\alpha,\alpha^{\prime}=x,y$ and $\alpha\neq\alpha^{\prime}$.

For the ground states, they can be formally diagonalized as $\vert g_{i}\rangle=\sum_{j=-1}^{1}C_{i,j}^{g}\vert\Phi_{A_{2};1,j}^{c}\rangle$ ($i=1,2,3$)
with eigenenergies $E_{i}^{g}$. And for the excited states, they can
be formally diagonalized in two subsets according to their polarizations
as $\vert e_{i}^{x}\rangle=\sum_{j=-1}^{1}C_{i,j}^{e}\vert\Phi_{E,x;1,j}^{c}\rangle$
and $\vert e_{i}^{y}\rangle=\sum_{j=-1}^{1}C_{i,j}^{e}\vert\Phi_{E,y;1,j}^{c}\rangle$ ($i=1,2,3$)
with degenerate eigenenergies $E_{i}^{e}$, where they share the same
coefficients due to the degeneracy.

According to Refs.~\cite{Jackson1999,Landau1995}, the constitutive relation reads
$\vec{D}=\epsilon_{0}\overleftrightarrow{\epsilon_{r}}\vec{E}=\epsilon_{D}\epsilon_{0}\vec{E}+\vec{P}$,
where $\vec{D}$ is the electric displacement,
$\epsilon_{0}$ is the electric permittivity of vacuum,
$\overleftrightarrow{\epsilon_{r}}$ and $\epsilon_{D}$ are, respectively,
the relative permittivity tensor of diamond with and without NV centers.
The polarization density can be calculated using linear response theory \cite{Kubo1985,Fang2016} as
\begin{eqnarray}
\vec{P} & = & -\frac{n_0}{\hbar }\mathrm{Re}\sum_{j,i,f}\rho_{i}\frac{\vec{d}^{(j)}_{if}(\vec{d}^{(j)}_{fi}\cdot\vec{E})}{\omega-\Delta^{(j)}_{fi}+i\gamma},
\label{eq:P}
\end{eqnarray}
where $\hbar$ is the Planck constant, $n_0=v_{0}^{-1}$ is the density
of the NV centers, $\rho_{i}$ is the probability of the initial state
$i$, $\omega$ is the frequency of the electric field $\vec{E}$, $\Delta^{(j)}_{fi}$
is the transition frequency of the $j$th NV center between the initial state $i$ and the
final state $f$, $\gamma^{-1}$ is the lifetime of the final state
$f$, and $\vec{d}^{(j)}_{if}$ is the transition matrix element of the electric
dipole of the $j$th NV center between the initial and final states.
Note that $i\neq f$ in Eq.~(\ref{eq:P}), and also in Eq.~(\ref{eq:Epsilon}).
In the above equation,
we did not explicitly discriminate the contributions from $\vert e_{i}^{x}\rangle$
and $\vert e_{i}^{y}\rangle$, as they only differ by the polarization
direction.

The relative permittivity tensor is \cite{SuppMat}
\begin{eqnarray}
\overleftrightarrow{\epsilon_{r}}\! & = & \! \epsilon_{D}\!-\!\mathrm{Re}\!\sum_{j,i,f,j_{1},j_{2}}\frac{C_{i,j_{1}}^{g*}(j)C_{f,j_{1}}^{e}(j)C_{f,j_{2}}^{e*}(j)C_{i,j_{2}}^{g}(j)}{3\hbar\epsilon_{0}v_{0}\left\{\omega-[E_{f}^{e}(j)-E_{i}^{g}(j)]+i\gamma\right\}}\nonumber \\
\! &  & \!\times(\vec{d}^{(j)}_{x}+\vec{d}^{(j)}_{y})(\vec{d}^{(j)}_{x}+\vec{d}^{(j)}_{y}),
 \label{eq:Epsilon}
\end{eqnarray}
where $\vec{d}^{(j)}_{x}$ and $\vec{d}^{(j)}_{y}$ are the components of the
transition dipole of the $j$th NV center. Clearly, there can be
nine possible negative permittivity components around the nine transition frequencies $\Delta^{(j)}_{fi}=E_{i}^{e}(j)-E_{f}^{g}(j)$ of the $j$th NV center.
However, if a static magnetic field is applied along the $z$-axis,
all transition frequencies would be correspondingly identical
for all four possible orientations \cite{Zou2014}.
Moreover, the relative permeability is not modified by the presence
of NV centers because the transition $^{3}A_{2}\rightleftharpoons{}^{3}E$
can only be induced by the electric-dipole couplings to the electromagnetic
field.

\begin{figure}
\includegraphics[bb=20 0 390 300,width=8.5cm]{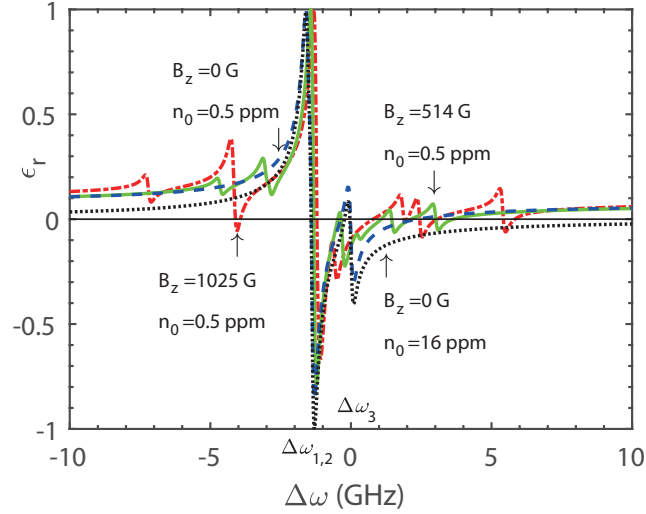}
\caption{(color online) The frequency dependence of the electric permittivity $\protect\overleftrightarrow{\epsilon_{r}}$
of diamond with NV centers for different values of the magnetic field $B$
and density of NV centers $n_{0}$: blue dashed line for $B_z=0$~G and $n_{0}=0.5$~ppm, green solid line for $B_z=514$~G and $n_{0}=0.5$~ppm, red dash-dotted line for $B_z=1025$~G and
$n_{0}=0.5$~ppm, black dotted line for $B_z=0$~G and $n_{0}=16$~ppm. Other
parameters are $d_{x}=d_{y}=11$~D \cite{Lenef1996}, $\gamma^{-1}=10$
ns \cite{Acosta2011}, $\epsilon_{D}=5.7$ \cite{Fontanella1977},
and $\mu_{D}=1-2.1\times10^{-5}$ \cite{Young1992}, $B_x=B_y=0$ G.
The thin black line $\protect\epsilon_{r}=0$ is just a guide to the eye. }
\label{fig:EpsilonVsB}
\end{figure}

In Fig.~\ref{fig:EpsilonVsB} we investigate the dependence of the permittivity
on the magnetic field and density of the NV centers.
Noticeably, two of the three components of the permittivity remain unchanged and
only one component is subtly modified by $\vec{B}$ due to the symmetry
and the special choice of $\vec{B}\parallel\vec{e_z}$ \cite{SuppMat}.
As a special case, we plot the modified component of the permittivity $\epsilon_r$
versus the frequency of the incident light for $B=0$.
When the magnetic field is absent,
in the manifold of the electronic ground state,
$\vert\Phi_{A_{2};1,\pm1}^{c}\rangle$ are degenerate and there is an energy gap $D_{\mathrm{gs}}$ between them and $\vert\Phi_{A_{2};1,0}^{c}\rangle$.
For the manifold of the electronic excited state,
because $\vert\Phi_{E,\alpha;1,\pm1}^{c}\rangle$ are degenerate,
there would be level anti-crossing due to the strain-related coupling $\xi$.
Since the electric-dipole induced transitions conserve the spin momentum \cite{Togan2010},
there could exist negative permittivity around three transition frequencies,
i.e. $\Delta\omega_{1,2}=D_{\mathrm{es}}^{\parallel}-D_{\mathrm{gs}}\pm\xi$ and $\Delta\omega_{3}=0$~GHz \cite{SuppMat}.
However, in the blue dashed curve of Fig.~\ref{fig:EpsilonVsB},
we can only observe two negative dips around
the transition frequencies $D_{\mathrm{es}}^{\parallel}-D_{\mathrm{gs}}=-1.46$~GHz and 0~GHz.
The first two dips merge into a single one as their widths, $\sim$~GHz, are
much larger than their separation, i.e. $2\xi=0.14$~GHz.
As the frequency of the incident light grows,
the modified component of the permittivity eventually increases to become positive at $\Delta\omega=2.37$~GHz. Therefore, for $B=0$ and $n_0=0.5$~ppm,
the frequency window for demonstrating negative refraction is roughly (-1.46,2.37)~GHz.
Because Eq.~(\ref{eq:Epsilon}) suggests negative permittivity to be around the transition frequencies, we hereafter explore the possibility of the negative refraction
beyond the above frequency domain by tuning magnetic field.
In the green solid curve of Fig.~\ref{fig:EpsilonVsB},
we plot the permittivity at the degenerate point of the excited states,
i.e., $B=514$~G. A new negative dip appears at $\Delta\omega=3.11$~GHz.
Interestingly, for the degenerate point of the ground states, i.e., $B=1025$~G,
in addition to the other two at $\Delta\omega=2.53$~GHz
and $\Delta\omega=5.51$~GHz on the right,
there is a new negative dip at $\Delta\omega=-4.06$~GHz
on the left hand side of $\Delta\omega_{1,2}$,
cf. red dash-dotted curve of Fig.~\ref{fig:EpsilonVsB}.
Meanwhile, the depth of the main dip at $\Delta\omega_{1,2}$ has been reduced
as compared to the case when $B=0$~G.
The increasing magnetic field does not only modify the transition frequencies,
but also redistributes electric dipoles among the eigenstates. In this regard,
by tuning the magnetic field, we can switch on/off the negative refraction on demand.
In Fig.~\ref{fig:EpsilonVsB}, there are only seven dips for the cases with $B=514$~G and $B=1025$~G, because there are two sets of degenerate eigenstates as shown in \cite{SuppMat}.
Furthermore, suggested by Eq.~(\ref{eq:Epsilon}),
the permittivity is also influenced by the density $n_{0}$ of the NV centers.
In the black dotted curve of Fig.~\ref{fig:EpsilonVsB},
this density is increased from $n_{0}=0.5$~ppm to $n_{0}=16$~ppm.
Compared to the blue dashed curve of Fig.~\ref{fig:EpsilonVsB},
the window of the negative refraction has been significantly broadened
because more NV centers can negatively respond to the applied magnetic field.
With this increased density, the negative dip at $\Delta\omega=-4.06$~GHz
can be more profound for $B=1025$~G.
Notice that in the numerical simulation we have not used the local field correction \cite{Jackson1999,Landau1995},
since the local field correction will not substantially change the center and
width of the negative refraction domain but will modify its magnitude \cite{Kastel2007-1,Kastel2007-2,Fang2016,SuppMat}.
Therefore, we have demonstrated negative refraction by the normalized $\epsilon_r$ in Fig.~\ref{fig:EpsilonVsB}.

\textit{Negative Refraction}.\textendash{}\textendash{}In Ref.~\cite{Veselago1968},
it has been shown that for a double-negative metamaterial there
can be negative refraction. However, in the NV centers of diamond,
because the electric permittivity tensor possesses
two different components, it is natural to ask whether
negative refraction can exist. Below, we will demonstrate that
negative refraction can indeed occur for a TH incident mode \cite{SuppMat}, cf.
Fig.~\ref{fig:scheme}(b).

According to Maxwell's equations \cite{Jackson1999,Landau1995},
$\nabla\times\vec{E} =-\frac{\partial}{\partial t}\mu_{D}\vec{H}$,
$\nabla\times\vec{H} =\frac{\partial}{\partial t}\overleftrightarrow{\epsilon}\vec{E}$,
where both the current density and the charge density vanish,
$\overleftrightarrow{\epsilon}=\epsilon_{0}\overleftrightarrow{\epsilon_{r}}$
is the permittivity of diamond with NV centers, and $\mu_{D}$ is the
permeability of pure diamond.

Assuming that the transmitted electric and magnetic fields are, respectively,
$\vec{E}_{t}(\vec{r},t) =(E_{tx}\hat{e}_{x}+E_{tz}\hat{e}_{z})\exp[i(\vec{k}_{t}\cdot\vec{r}-\omega t)]$,
$\vec{H}_{t}(\vec{r},t) =H_{ty}\hat{e}_{y}\exp[i(\vec{k}_{t}\cdot\vec{r}-\omega t)]$,
we have
\begin{equation}
(\nabla\times\nabla\times\overleftrightarrow{I}-\mu_{0}\omega^{2}\overleftrightarrow{\epsilon})\vec{E}_{t}=0,\label{eq:Eigen}
\end{equation}
where $\overleftrightarrow{I}$ is the identity dyadic. For nontrivial
solutions, the dispersion relation for the extraordinary mode reads
\begin{align}
\epsilon_{x}k_{tx}^{2}+\epsilon_{z}k_{tz}^{2} & =\mu_{0}\omega^{2}\epsilon_{x}\epsilon_{z},\label{eq:ExtraOrDispersion}
\end{align}
assuming $k_{y}=0$. Such a dispersion relation
for the extraordinary mode is hyperbolic or indefinite because $\epsilon_{x}\epsilon_{z}<0$.

According to the boundary condition~\cite{Jackson1999},
the tangential components of the wavevector across the interface should
be equal, i.e.,
$k_{tz} =k_{iz}>0$,
$k_{tx} =k_{ix}$.
By inserting Eq.~(\ref{eq:ExtraOrDispersion}) into Eq.~(\ref{eq:Eigen}),
we obtain the relation between $E_{tx}$ and $E_{tz}$ as
$\epsilon_{x}k_{tx}E_{tx}+\epsilon_{z}k_{tz}E_{tz}=0$.
By Maxwell equation, we have
\begin{align}
\vec{H} =-\frac{\omega\epsilon_{z}E_{tz}}{k_{tx}}\hat{e}_{y}\exp\left[i\left(\vec{k}_{t}\cdot\vec{r}-\omega t\right)\right].
\end{align}
The time-averaged Poynting vector reads \cite{Jackson1999}
$\vec{S}_{t}=\frac{1}{2}\mathrm{Re}(\vec{E}_{t}\times\vec{H}_{t}^{*})$,
with the components being
$S_{tx} =\frac{\omega\epsilon_{z}}{2k_{tx}}E_{tz}^{2}$,
$S_{tz} =\frac{\epsilon_{x}\omega E_{tx}^{2}}{2k_{tz}}<0$,
because $\epsilon_{x}<0$ and $\omega,k_{tz}>0$. In order to transmit
energy from the interface into the medium, $S_{tx}$ should be negative
and thus $k_{tx}<0$ as $\omega,\epsilon_{z}>0$. Together with Eq.~(\ref{eq:ExtraOrDispersion}),
we have
\begin{align}
k_{tx} & =-k_{i}\sqrt{\frac{\epsilon_{z}}{\epsilon_{0}}\left(1-\frac{\epsilon_{0}}{\epsilon_{x}}\sin^{2}\theta\right)},
\end{align}
where
$k_{i}^{2}=\mu_{0}\epsilon_{0}\omega^{2}$.
Because $S_{tx},S_{tz}<0$, we have proven that for
a uniaxial crystal with hyperbolic dispersion, the negative
refraction exists for a TH incident field.

\textit{Experimental Feasibility}.\textendash{}\textendash{}For zero magnetic field,
the Hamiltonians of the electronic ground and excited states are further simplified as $H_{\mathrm{gs}}=D_{\mathrm{gs}}\sum_{m_{z}=\pm1}\vert\Phi_{A_{2};1,m_{z}}^{c}\rangle\langle\Phi_{A_{2};1,m_{z}}^{c}\vert$
and $H_{\mathrm{es}}\simeq\sum_{\alpha=x,y}\sum_{m_{z}=\pm1}D_{\mathrm{es}}^{\parallel}\vert\Phi_{E,\alpha;1,m_{z}}^{c}\rangle\langle\Phi_{E,\alpha;1,m_{z}}^{c}\vert$ \cite{SuppMat},
where we have omitted the strain-related coupling.

The transition electric dipole has been estimated as $11$~D \cite{Lenef1996}.
For simplicity, the orientations of all NV centers are assumed
to be along the $z$-axis. Thus, all matrix
elements of the transition electric dipole are equal to
$\vec{d}_{if}=\langle\Phi_{A_{2};1,m_{z}}^{c}\vert\vec{d}\vert\Phi_{E,\alpha;1,m_{z}}^{c}\rangle=11(\hat{e}_{x}+\hat{e}_{y}) $~D.
Initially, the NV center is in the thermal state $\rho(0)=\frac{1}{3}\sum_{m_{z}=\pm1}\vert\Phi_{A_{2};S,m_{s}}^{c}\rangle\langle\Phi_{A_{2};S,m_{s}}^{c}\vert$.
Therefore,
$\sum_{i,f}\vec{d}_{if}\vec{d}_{fi}= \frac{484}{3}(\hat{e}_{x}\hat{e}_{x}+\hat{e}_{y}\hat{e}_{y}+\hat{e}_{x}\hat{e}_{y}+\hat{e}_{y}\hat{e}_{x}) \; \mathrm{D}^{2}$, and
\begin{eqnarray}
\vec{P} & = & -2\zeta\gamma\epsilon_0\mathrm{Re}\left[\frac{(\hat{e}_{x}\hat{e}_{x}+\hat{e}_{y}\hat{e}_{y}+\hat{e}_{x}\hat{e}_{y}+\hat{e}_{y}\hat{e}_{x})\vec{E}}{\omega-\Delta_{fi}+i\gamma}\right],
\end{eqnarray}
where $\zeta = \frac{242n_{0}\;\mathrm{D}^{2}}{9\hbar\gamma\epsilon_0}$. The three
principal components of the relative permittivity are, respectively,
\begin{equation}
\epsilon^{(1)}_r=\epsilon_{D}-\frac{2\zeta\gamma(\omega-\Delta_{fi})}{(\omega-\Delta_{fi})^2+\gamma^2},
\end{equation}
$\epsilon^{(2)}_r=\epsilon^{(3)}_r=\epsilon_{D}$.
When the frequency of the incident field is $\omega=\Delta_{fi}+\gamma$,
one principal component can be negative if
$n_{0}>n^c_{0}=1.77\times10^{21} \; \mathrm{m}^{-3}$,
while the other principal components remain positive.
Because two carbon atoms occupy a volume $v=(1.78\times10^{-10})^3 \; \mathrm{m}^{-3}$,
the minimum density of the NV centers to demonstrate negative refraction is
\begin{equation}
\frac{1}{2}vn^c_0=5.00 \; \mathrm{ppb},
\end{equation}
which is feasible in experimental fabrication, e.g. 16~ppm
\cite{Jarmola2012}. In addition, as proven in \cite{SuppMat},
the negative component of permittivity appears in the $z$-axis,
because of $\vec{B}\parallel\vec{e}_z$
and the symmetry of four possible orientations of the NV centers.

\textit{Conclusions}.\textendash{}\textendash{}In this work,
we proposed a new approach to realize hyperbolic metamaterial
using diamond with NV centers in the optical frequency regime.
Because of the long lifetime of the excited states of the NV centers,
this hyperbolic metamaterial manifests an intriguing window for negative refraction.
By varying the applied magnetic field to
tune the energy spectra of both ground and excited states,
the frequency of the negative refraction can be tuned in a wide range.
Note that it is difficult to fabricate classical metamaterials
working in optical-frequency domain,
because the sizes of the elements therein are sub-micron.
The NV-metamaterials proposed here solve this problem.
Because this NV-metamaterial can be used in subwavelength imaging,
spontaneous emission enhancement, heat transport and acoustics,
analogue cosmology, and lifetime engineering,
our proposal bridges the gap between NV centers and metamaterials.

\begin{acknowledgments}
We thank stimulating discussion with Zhou Li and K. Y. Bliokh.
This work was supported by the MURI Center for Dynamic Magneto-Optics via the
AFOSR Award No.~FA9550-14-1-0040, the Japan Society
for the Promotion of Science (KAKENHI), the IMPACT program of JST,
JSPS-RFBR grant No.~17-52-50023, CREST grant No.~JPMJCR1676,
and RIKEN-AIST Challenge Research Fund. C.P.S. was supported by
NSFC under Grant No.~11421063 and No.~11534002, NSAF under Grant No.~U1530401.
Q.A. was partially supported by NSFC under Grant No.~11505007.
\end{acknowledgments}

\include{SuppleMaterial09}

\end{document}

%% file: SuppleMaterial09.tex
\title{Supplemental Material for ``Tunable Quantum Hyperbolic Metamaterial
Using Nitrogen-Vacancy Centers in Diamond''}

\author{Qing Ai}

\address{CEMS, RIKEN, Wako-shi, Saitama 351-0198, Japan}

\address{Department of Physics, Applied Optics Beijing Area Major Laboratory,
Beijing Normal University, Beijing 100875, China}

\author{Peng-Bo Li}

\address{CEMS, RIKEN, Wako-shi, Saitama 351-0198, Japan}

\address{Department of Applied Physics, Xi\textquoteright an Jiaotong University,
Xi\textquoteright an 710049, China}

\author{Wei Qin}

\address{CEMS, RIKEN, Wako-shi, Saitama 351-0198, Japan}

\address{Quantum Physics and Quantum Information Division, Beijing Computational
Science Research Center, Beijing 100193, China}

\author{C. P. Sun}

\address{Beijing Computational Science Research Center \& Graduate School
of Chinese Academy of Engineering Physics, Beijing 100084, China}

\author{Franco Nori}

\address{CEMS, RIKEN, Wako-shi, Saitama 351-0198, Japan}

\address{Department of Physics, The University of Michigan, Ann Arbor, Michigan
48109-1040, USA}
%

\maketitle

\setcounter{equation}{0} 
\setcounter{figure}{0} 
\setcounter{table}{0} 
\setcounter{page}{1} 
\setcounter{section}{0} 
\makeatletter 
\renewcommand{\theequation}{S\arabic{equation}} 
\renewcommand{\thefigure}{S\arabic{figure}} 
\renewcommand{\thepage}{S\arabic{page}} 
\renewcommand{\thesection}{S\Roman{section}} 
\renewcommand{\bibnumfmt}[1]{[S#1]} 
\renewcommand{\citenumfont}[1]{S#1}

\section{Model}
\label{sec:Model}

The Hamiltonian of an NV center in its electronic \textit{ground}
state is \cite{Doherty2013}
\begin{eqnarray}
H_{\mathrm{gs}} & = & D_{\mathrm{gs}}\left[S_{z}^{2}-\frac{1}{3}S(S+1)\right]+A_{\mathrm{gs}}^{\parallel}S_{z}I_{z}+A_{\mathrm{gs}}^{\perp}(S_{x}I_{x}+S_{y}I_{y})+P_{\mathrm{gs}}\left[I_{z}^{2}-\frac{1}{3}I(I+1)\right]\nonumber \\
 &  & +\mu_{B}g_{\mathrm{gs}}^{\parallel}B_{z}S_{z}+\mu_{B}g_{\mathrm{gs}}^{\perp}(B_{x}S_{x}+B_{y}S_{y})+\mu_{N}g_{N}\vec{B}\cdot\vec{I}\nonumber \\
 &  & +d_{\mathrm{gs}}^{\parallel}(E_{z}+\delta_{z})\left[S_{z}^{2}-\frac{1}{3}S(S+1)\right]+d_{\mathrm{gs}}^{\perp}(E_{x}+\delta_{x})(S_{y}^{2}-S_{x}^{2})+d_{\mathrm{gs}}^{\perp}(E_{y}+\delta_{y})(S_{x}S_{y}+S_{y}S_{x}).
\end{eqnarray}
Here, $D_{\mathrm{gs}}=2.88$ GHz is the zero-field splitting of the
electronic ground state; $A_{\mathrm{gs}}^{\parallel}$ and $A_{\mathrm{gs}}^{\perp}$
are the axial and non-axial components of hyperfine interaction tensor
of the electronic ground state; $I_{x}$, $I_{y}$, $I_{z}$ are the
spin operators of the nuclear spin; $P_{\mathrm{gs}}$ is the nuclear
electric quadruple parameter of the electronic ground state; $\mu_{B}$
and $\mu_{N}$ are the Bohr magneton and nuclear magneton respectively;
$g_{\mathrm{gs}}^{\parallel}\simeq g_{\mathrm{gs}}^{\perp}=g_{\mathrm{gs}}\simeq2$
and $g_{N}$ are respectively the $g$-factors of electronic ground
state and nuclear spin; $d_{\mathrm{gs}}^{\parallel}=3.377\times10^{-5}$
D and $d_{\mathrm{gs}}^{\perp}=6.9525\times10^{-7}$ D are the components
of electric dipole moment of the electronic ground state; $\vec{E}$,
$\vec{B}$, and $\vec{\delta}$ are the electric, magnetic and strain
fields respectively. The electron spin operators in the basis \{$\vert+\rangle$,$\vert0\rangle$,$\vert-\rangle$\}
are
\begin{align}
S_{x} & =\frac{1}{\sqrt{2}}\begin{pmatrix}0 & 1 & 0\\
1 & 0 & 1\\
0 & 1 & 0
\end{pmatrix},\\
S_{y} & =\frac{i}{\sqrt{2}}\begin{pmatrix}0 & -1 & 0\\
1 & 0 & -1\\
0 & 1 & 0
\end{pmatrix},\\
S_{z} & =\begin{pmatrix}1 & 0 & 0\\
0 & 0 & 0\\
0 & 0 & -1
\end{pmatrix}.
\end{align}
When there is no nuclear spin, electric and strain fields, the Hamiltonian
of the electronic ground state is simplified as
\begin{eqnarray}
H_{\mathrm{gs}} & = & D_{\mathrm{gs}}S_{z}^{2}+\mu_{B}g_{\mathrm{gs}}^{\parallel}B_{z}S_{z}+\mu_{B}g_{\mathrm{gs}}^{\perp}(B_{x}S_{x}+B_{y}S_{y})\nonumber \\
 & = & \begin{pmatrix}D_{\mathrm{gs}}+\mu_{B}g_{\mathrm{gs}}B_{z} & \frac{1}{\sqrt{2}}\mu_{B}g_{\mathrm{gs}}(B_{x}-iB_{y}) & 0\\
\frac{1}{\sqrt{2}}\mu_{B}g_{\mathrm{gs}}(B_{x}+iB_{y}) & 0 & \frac{1}{\sqrt{2}}\mu_{B}g_{\mathrm{gs}}(B_{x}-iB_{y})\\
0 & \frac{1}{\sqrt{2}}\mu_{B}g_{\mathrm{gs}}(B_{x}+iB_{y}) & D_{\mathrm{gs}}-\mu_{B}g_{\mathrm{gs}}^{\parallel}B_{z}
\end{pmatrix}.
\end{eqnarray}

At room temperature, the Hamiltonian of an NV center in the electronic
\textit{excited} state is \cite{Doherty2013}
\begin{eqnarray}
H_{\mathrm{es}} & = & D_{\mathrm{es}}^{\parallel}\left[S_{z}^{2}-\frac{1}{3}S(S+1)\right]+A_{\mathrm{es}}^{\parallel}S_{z}I_{z}+A_{\mathrm{es}}^{\perp}(S_{x}I_{x}+S_{y}I_{y})+P_{\mathrm{es}}\left[I_{z}^{2}-\frac{1}{3}I(I+1)\right]\nonumber \\
 &  & +\mu_{B}g_{\mathrm{es}}^{\mathrm{RT}}\vec{B}\cdot\vec{S}+d_{\mathrm{es}}^{\parallel}(E_{z}+\delta_{z})\left[S_{z}^{2}-\frac{1}{3}S(S+1)\right]+\xi(S_{y}^{2}-S_{x}^{2}),
\end{eqnarray}
where $D_{\mathrm{es}}^{\parallel}=1.42$ GHz is the zero-field splitting
of the electronic excited state; $A_{\mathrm{es}}^{\parallel}$ and
$A_{\mathrm{es}}^{\perp}$ are the axial and non-axial components
of hyperfine interaction tensor of the electronic excited state; $P_{\mathrm{es}}$
is the nuclear electric quadruple parameter of the electronic excited
state; $g_{\mathrm{es}}^{\mathrm{RT}}\simeq2.01$ is the $g$-factor of electronic
spin of excited state at the room temperature; $d_{\mathrm{es}}^{\parallel}=1.192$
D is the electric dipole moment of the excited state; $\xi=70$ MHz
is the strain-related coupling. When there is no nuclear spin, electric
and strain fields, the Hamiltonian of the electronic excited state
is simplified as
\begin{eqnarray}
H_{\mathrm{es}} & = & D_{\mathrm{es}}^{\parallel}S_{z}^{2}+\mu_{B}g_{\mathrm{es}}^{\mathrm{RT}}\vec{B}\cdot\vec{S}+\xi(S_{y}^{2}-S_{x}^{2})\nonumber \\
 & = & \begin{pmatrix}D_{\mathrm{es}}^{\parallel}+\mu_{B}g_{\mathrm{es}}^{\mathrm{RT}}B_{z} & \frac{1}{\sqrt{2}}\mu_{B}g_{\mathrm{es}}^{\mathrm{RT}}(B_{x}-iB_{y}) & -\xi\\
\frac{1}{\sqrt{2}}\mu_{B}g_{\mathrm{es}}^{\mathrm{RT}}(B_{x}+iB_{y}) & 0 & \frac{1}{\sqrt{2}}\mu_{B}g_{\mathrm{es}}^{\mathrm{RT}}(B_{x}-iB_{y})\\
-\xi & \frac{1}{\sqrt{2}}\mu_{B}g_{\mathrm{es}}^{\mathrm{RT}}(B_{x}+iB_{y}) & D_{\mathrm{es}}^{\parallel}-\mu_{B}g_{\mathrm{es}}^{\mathrm{RT}}B_{z}
\end{pmatrix}.
\end{eqnarray}

\section{Linear Response Theory}
\label{sec:LRT}

In order to simulate the electromagnetic response of the diamond with
NV centers in the presence of applied fields, we can employ the linear-response
theory \cite{Kubo1985} to calculate the electric permittivity and
magnetic permeability. When there is an electric field applied, the
NV center is polarized as
\begin{equation}
\langle\vec{d}\,\rangle=\int\frac{d\omega}{2\pi}S(\omega)\vec{E}(\omega)e^{-i\omega t},
\end{equation}
where the Fourier transform of the time-dependent electric field with
amplitude $\vec{E}{}_{0}$ and frequency $\omega$
\begin{equation}
\vec{E}(t)=\vec{E}{}_{0}\cos\omega t
\end{equation}
is
\begin{align}
\vec{E}(\omega) & =\int_{-\infty}^{\infty}dt\;\vec{E}(t)e^{i\omega t},\\
S(\omega) & =-J(\omega)-J^{\ast}(-\omega).
\end{align}
Here, $J(\omega)$ is the dipole-dipole correlation function,

\begin{equation}
J(\omega)=-i\int_{0}^{\infty}dt\;\mathrm{\textrm{Tr}}[\vec{d}(t)\vec{d}\rho_{0}]e^{i\omega t},
\end{equation}
where the initial state of the NV center is
\begin{equation}
\rho_{0}=\sum_{i}\rho_i\vert k_{i}\rangle\langle k_{i}\vert
\end{equation}
with $\sum_{i}\rho_i=1$.

The electric dipole in the Heisenberg picture is
\begin{equation}
\vec{d}(t)=\exp(iH^{\dagger}t/\hbar)\vec{d}\exp(-iHt/\hbar),
\end{equation}
where
\begin{equation}
H=H_{\mathrm{es}}\otimes\vert e\rangle\langle e\vert+H_{\mathrm{gs}}\otimes\vert g\rangle\langle g\vert,
\end{equation}
with $\vert g\rangle$ ($\vert e\rangle$) being the electronic ground (excited) state.
Because the Fourier transform of the electric field is
\begin{align}
\vec{E}(\omega_{1}) =\int_{-\infty}^{\infty}dt\;\vec{E}_{0}\cos\omega te^{i\omega_{1}t} =\pi\vec{E}_{0}[\delta(\omega_{1}+\omega)+\delta(\omega_{1}-\omega)],
\end{align}
the electric dipole of the NV center in the applied electric field is
\begin{eqnarray}
\langle\vec{d}\,\rangle = \int\frac{d\omega_{1}}{2\pi}S(\omega_{1})e^{-i\omega_{1}t}\pi\vec{E}_{0}[\delta(\omega_{1}+\omega)+\delta(\omega_{1}-\omega)]
 = -\vec{E}_{0}\mathrm{Re}\{\left[J(\omega)+J^{\ast}(-\omega)\right]e^{-i\omega t}\},
\end{eqnarray}
where
\begin{eqnarray}
J(\omega) & = & -i\int_{0}^{\infty}dt\;e^{i\omega t}\sum_{i}\rho_i\langle k_{i}\vert\vec{d}(t)\vec{d}\vert k_{i}\rangle\nonumber \\
 & = & -i\int_{0}^{\infty}dt\;e^{i\omega t}\sum_{i}\rho_i\langle k_{i}\vert e^{iH^{\dagger}t/\hbar}\vec{d}e^{-iHt/\hbar}\vec{d}\vert k_{i}\rangle\nonumber \\
 & = & -i\int_{0}^{\infty}dt\;e^{i\omega t}\sum_{i,k_{1},k_{2},k_{3}}\rho_i\langle k_{i}\vert e^{iH^{\dagger}t/\hbar}\left|k_{1}\right\rangle \left\langle k_{1}\right|\vec{d}\left|k_{2}\right\rangle \left\langle k_{2}\right|e^{-iHt/\hbar}\left|k_{3}\right\rangle \left\langle k_{3}\right|\vec{d}\left|k_{i}\right\rangle \nonumber \\
 & = & -i\int_{0}^{\infty}dt\;e^{i\omega t}\sum_{i,f}\rho_ie^{i\left(H_{k_{i}}^{\dagger}-H_{k_{f}}\right)t/\hbar}\vec{d}_{k_{i},k_{f}}\vec{d}_{k_{f},k_{i}}\nonumber \\
 & = & -i\int_{0}^{\infty}dt\sum_{i,f}\rho_ie^{i\left(\omega-\Delta_{k_{f}k_{i}}+i\gamma\right)t}\vec{d}_{k_{i},k_{f}}\vec{d}_{k_{f},k_{i}}\nonumber \\
 & = & \sum_{i,f}\rho_i\frac{\vec{d}_{k_{i},k_{f}}\vec{d}_{k_{f},k_{i}}}{\omega-\Delta_{k_{f}k_{i}}+i\gamma},
\end{eqnarray}
where in the sum the final state should be different from the initial
state, i.e., $i\neq f$,
\begin{eqnarray}
H_{k_{i}} & = & \left\langle k_{i}\right|H\left|k_{i}\right\rangle =E_{k_{i}}-\frac{i}{2}\gamma
\end{eqnarray}
with $-i\gamma/2$ being phenomenologically introduced for the decay
of the excited state. Here, $\Delta_{k_{f}k_{i}}=E_{k_{f}}-E_{k_{i}}$
is the transition energy between the initial state $\left|k_{i}\right\rangle $
and the final state $\left|k_{f}\right\rangle $. Therefore, the induced
electric dipole can be rewritten as

\begin{eqnarray}
\langle\vec{d}\,\rangle & = & -\vec{E}_{0}\;\mathrm{Re}\left\{ \sum_{k_{i},k_{f}}\rho_i\left[\frac{\vec{d}_{k_{i},k_{f}}\vec{d}_{k_{f},k_{i}}}{(\omega-\Delta_{k_{f}k_{i}}+i\gamma)}-\frac{\vec{d}_{k_{i},k_{f}}\vec{d}_{k_{f},k_{i}}}{(\omega+\Delta_{k_{f}k_{i}}+i\gamma)}\right]e^{-i\omega t}\right\} .
\end{eqnarray}
Because of the rotating-wave approximation \cite{Ai2010}, the second
term of the above equation should be neglected, i.e.
\begin{eqnarray}
\langle\vec{d}\,\rangle & \approx & -\vec{E}_{0}\;\mathrm{Re}\left[\sum_{k_{i},k_{f}}\frac{\rho_i\vec{d}_{k_{i},k_{f}}\vec{d}_{k_{f},k_{i}}}{\omega-\Delta_{k_{f}k_{i}}+i\gamma}e^{-i\omega t}\right].
\end{eqnarray}
Assuming that all NV centers are identical, the polarization density
reads
\begin{equation}
\vec{P}=\frac{n_{0}}{\hbar}\langle\vec{d}\,\rangle,
\end{equation}
where $n_{0}$ is the number density of the NV centers in diamond.

\section{Lorentz Local Field Theory}

According to Ref.~\cite{Jackson1999}, in closely-packed molecules
the polarization of neighboring molecules gives rise to an internal
field $\vec{E_{i}}$ at any molecule, in addition to the external
field $\vec{E}$. The internal field is
\begin{equation}
\vec{E}_{i}=\vec{E}_{\mathrm{near}}-\vec{E}_{\mathrm{mean}},
\end{equation}
where $\vec{E}_{\mathrm{near}}$ is the actual contribution from the
molecules close to the given molecule, and $\vec{E}_{\mathrm{mean}}$ is the contribution
from those molecules treated in an average continuum. As proven in
Ref.~\cite{Jackson1999}, in any crystal structure $\vec{E}_{\mathrm{near}}=0$
due to symmetry, and thus $\vec{E}_{i}=-\vec{E}_{\mathrm{mean}}$.

By dipole approximation and assuming no net charge in the volume $V$,
the mean-field contribution is \cite{Jackson1999}
\begin{equation}
\vec{E}_{\mathrm{mean}}=-(\epsilon_D-1)\vec{E}-\frac{1}{3V\epsilon_0}\sum_l\vec{p}_l,
\end{equation}
where $\epsilon_{D}=5.7$ \cite{Fontanella1977} is relative permittivity of diamond,
and the second term is summed over all induced molecular electric dipole moments
$\vec{p}_l$ within the volume. Under the weak field approximation,
the induced dipole moment is
\begin{equation}
\vec{p}_l=\epsilon_0\gamma_{\mathrm{mol}}(\vec{E}+\vec{E}_i),
\end{equation}
where $\gamma_{\mathrm{mol}}$ is generally a second-order tensor.
Since $\vec{E}_{i}=(\epsilon_D-1)\vec{E}+\vec{P}/(3\epsilon_{0})$ \cite{Jackson1999,Marques2008},
the polarization $\vec{P}\equiv\sum_l\vec{p}_{l}/V=n_{0}\vec{p}_{l}$ reads
\begin{eqnarray}
\vec{P} = n_{0}\epsilon_{0}\gamma_{\mathrm{mol}}(\vec{E}+\vec{E}_{i}) = n_{0}\epsilon_{0}\gamma_{\mathrm{mol}}(\epsilon_D\vec{E}+\frac{\vec{P}}{3\epsilon_{0}}).
\end{eqnarray}
Furthermore, using $\vec{P}=\epsilon_{0}\chi_{e}\vec{E}$
\cite{Jackson1999,Marques2008}, we have
\begin{equation}
\epsilon_{0}\chi_{e}\vec{E}=n_{0}\epsilon_{0}\gamma_{\mathrm{mol}}(\epsilon_D+\frac{\chi_{e}}{3})\vec{E},
\end{equation}
leading to
\begin{equation}
\chi_{e}=\frac{n_{0}\gamma_{\mathrm{mol}}\epsilon_D}{1-\frac{1}{3}n_{0}\gamma_{\mathrm{mol}}}.\label{eq:3}
\end{equation}

As proven in Sec.~\ref{sec:Exp}, in diamond with NV centers, $\gamma_{\mathrm{mol}}$ is of the form
\begin{equation}
\gamma_{\mathrm{mol}}=\begin{pmatrix}\eta(\omega)/n_0 & 0 & 0\\
0 & 0 & 0\\
0 & 0 & 0
\end{pmatrix},
\end{equation}
where $\eta(\omega)$ is the second part of Eq.~(11) in the main text.
Therefore, the electric susceptibility is
\begin{equation}
\chi_{e}=\begin{pmatrix}\frac{3\epsilon_D\eta}{3-\eta} & 0 & 0\\
0 & 0 & 0\\
0 & 0 & 0
\end{pmatrix},
\end{equation}
and the relative permittivity is
\begin{equation}
\overrightarrow{\epsilon_{r}}=\begin{pmatrix}\epsilon_{D}(1+\frac{3\eta}{3-\eta}) & 0 & 0\\
0 & \epsilon_{D} & 0\\
0 & 0 & \epsilon_{D}
\end{pmatrix}.
\end{equation}
The window of the negative refraction is determined by the two solutions to the equation
\begin{equation}
1+\frac{3\eta}{3-\eta}=0,\textrm{or equivalently,}\;\frac{3}{2}+\eta=0,
\end{equation}
which is not qualitatively different from the equation for the case
without the local field correction
\begin{equation}
\epsilon_{D}+\eta=0.
\end{equation}
As a result,
both the center and width of the negative refraction domain are not
substantially modified by the Lorentz local field correction.
This is in consistent with the numerical results in Refs.~\cite{Kastel2007-1,Kastel2007-2,Fang2016}.

\section{Selection Rules of Optical Transitions}
\label{sec:SROT}

According to Ref.~\cite{Doherty2011}, there are four outer electrons
distributed in the $a_{1}$, $e_{x}$ and $e_{y}$ levels, i.e., $a_{1}^{2}e^{2}$.
On account of the spin degree of freedom, the electronic ground states
can be written in the second quantization form, $\vert a_{1}\bar{a}_{1}e_{x}\bar{e}_{x}e_{y}\bar{e}_{y}\rangle$
with an overbar denoting spin-down, as
\begin{align}
\vert\Phi_{A_{2};1,0}^{c}\rangle & =\frac{1}{\sqrt{2}}(\vert111001\rangle+\vert110110\rangle),\\
\vert\Phi_{A_{2};1,1}^{c}\rangle & =\vert111010\rangle,\\
\vert\Phi_{A_{2};1,-1}^{c}\rangle & =\vert110101\rangle,
\end{align}
where the superscript $c$ means configuration, the subscripts are
ordered as $j,k;S,m_{s}$ with $j$ being the irreducible representation,
$k$ being the row of irreducible representation, $S$ being the total
spin and $m_{s}$ being the spin projection along the symmetry axis
of the NV. The six first excited states, i.e., $a_{1}e^{3}$, are respectively
\begin{align}
\vert\Phi_{E,x;1,0}^{c}\rangle & =\frac{1}{\sqrt{2}}(\vert100111\rangle+\vert011011\rangle),\\
\vert\Phi_{E,y;1,0}^{c}\rangle & =\frac{1}{\sqrt{2}}(\vert101101\rangle+\vert011110\rangle),\\
\vert\Phi_{E,x;1,1}^{c}\rangle & =\vert101011\rangle,\\
\vert\Phi_{E,y;1,1}^{c}\rangle & =\vert101110\rangle,\\
\vert\Phi_{E,x;1,-1}^{c}\rangle & =\vert010111\rangle,\\
\vert\Phi_{E,y;1,-1}^{c}\rangle & =\vert011101\rangle.
\end{align}
By comparing the above states, we notice that there is one electron
transiting from the $a_{1}$ orbital to the $e$ orbital. In the absence
of spin-orbital coupling, on account of conservation of spin and total
angular momentum \cite{Togan2010}, the following transitions are
allowed by the electric dipole coupling
\begin{eqnarray}
\vert\Phi_{A_{2};1,0}^{c}\rangle & \rightleftharpoons & \vert\Phi_{E,x;1,0}^{c}\rangle,\vert\Phi_{E,y;1,0}^{c}\rangle,\\
\vert\Phi_{A_{2};1,1}^{c}\rangle & \rightleftharpoons & \vert\Phi_{E,x;1,1}^{c}\rangle,\vert\Phi_{E,y;1,1}^{c}\rangle,\\
\vert\Phi_{A_{2};1,-1}^{c}\rangle & \rightleftharpoons & \vert\Phi_{E,x;1,-1}^{c}\rangle,\vert\Phi_{E,y;1,-1}^{c}\rangle.
\end{eqnarray}
The non-zero transition matrix elements of the position vector $\vec{r}=x\hat{e}_{x}+y\hat{e}_{y}+z\hat{e}_{z}$
($\hat{e}_{i}$ unit vector along direction $i=x,y,z$) are listed as \cite{Maze2011}
\begin{align}
\langle a_{1}\vert x\vert e_{x}\rangle & \neq0,\\
\langle a_{1}\vert y\vert e_{y}\rangle & \neq0,\\
\langle e_{y}\vert y\vert e_{x}\rangle & =\langle e_{y}\vert x\vert e_{y}\rangle=\langle e_{x}\vert y\vert e_{y}\rangle=-\langle e_{x}\vert x\vert e_{x}\rangle\neq0.
\end{align}
Therefore, we have
\begin{align}
\langle\Phi_{A_{2};1,0}^{c}\vert x\vert\Phi_{E,x;1,0}^{c}\rangle & =\frac{1}{\sqrt{2}}(\langle111001\vert+\langle110110\vert)x\frac{1}{\sqrt{2}}(\vert100111\rangle+\vert011011\rangle)\nonumber \\
 & =\frac{1}{2}\langle a_{1}\vert x\vert e_{y}\rangle+\frac{1}{2}\langle\bar{a}_{1}\vert x\vert\bar{e}_{y}\rangle=0,\\
\langle\Phi_{A_{2};1,0}^{c}\vert y\vert\Phi_{E,x;1,0}^{c}\rangle & =\frac{1}{\sqrt{2}}(\langle111001\vert+\langle110110\vert)y\frac{1}{\sqrt{2}}(\vert100111\rangle+\vert011011\rangle)\nonumber \\
 & =\frac{1}{2}\langle a_{1}\vert y\vert e_{y}\rangle+\frac{1}{2}\langle\bar{a}_{1}\vert y\vert\bar{e}_{y}\rangle=\langle a_{1}\vert y\vert e_{y}\rangle,\\
\langle\Phi_{A_{2};1,0}^{c}\vert x\vert\Phi_{E,y;1,0}^{c}\rangle & =\frac{1}{\sqrt{2}}(\langle111001\vert+\langle110110\vert)x\frac{1}{\sqrt{2}}(\vert101101\rangle+\vert011110\rangle)\nonumber \\
 & =\frac{1}{2}\langle\bar{a}_{1}\vert x\vert\bar{e}_{x}\rangle+\frac{1}{2}\langle a_{1}\vert x\vert e_{x}\rangle=\langle a_{1}\vert x\vert e_{x}\rangle,\\
\langle\Phi_{A_{2};1,0}^{c}\vert y\vert\Phi_{E,y;1,0}^{c}\rangle & =\frac{1}{\sqrt{2}}(\langle111001\vert+\langle110110\vert)y\frac{1}{\sqrt{2}}(\vert101101\rangle+\vert011110\rangle)\nonumber \\
 & =\frac{1}{2}\langle\bar{a}_{1}\vert y\vert\bar{e}_{x}\rangle+\frac{1}{2}\langle a_{1}\vert y\vert e_{x}\rangle=0,
\end{align}
\begin{align}
\langle\Phi_{A_{2};1,1}^{c}\vert x\vert\Phi_{E,x;1,1}^{c}\rangle & =\langle111010\vert x\vert101011\rangle =\langle\bar{a}_{1}\vert x\vert\bar{e}_{y}\rangle=0,\\
\langle\Phi_{A_{2};1,1}^{c}\vert y\vert\Phi_{E,x;1,1}^{c}\rangle & =\langle111010\vert y\vert101011\rangle =\langle\bar{a}_{1}\vert y\vert\bar{e}_{y}\rangle,\\
\langle\Phi_{A_{2};1,1}^{c}\vert x\vert\Phi_{E,y;1,1}^{c}\rangle & =\langle111010\vert x\vert101110\rangle =\langle\bar{a}_{1}\vert x\vert\bar{e}_{x}\rangle,\\
\langle\Phi_{A_{2};1,1}^{c}\vert y\vert\Phi_{E,y;1,1}^{c}\rangle & =\langle111010\vert y\vert101110\rangle =\langle\bar{a}_{1}\vert y\vert\bar{e}_{x}\rangle=0,\\
\langle\Phi_{A_{2};1,-1}^{c}\vert x\vert\Phi_{E,x;1,-1}^{c}\rangle & =\langle110101\vert x\vert010111\rangle =\langle a_{1}\vert x\vert e_{y}\rangle=0,\\
\langle\Phi_{A_{2};1,-1}^{c}\vert y\vert\Phi_{E,x;1,-1}^{c}\rangle & =\langle110101\vert y\vert010111\rangle =\langle a_{1}\vert y\vert e_{y}\rangle,\\
\langle\Phi_{A_{2};1,-1}^{c}\vert x\vert\Phi_{E,y;1,-1}^{c}\rangle & =\langle110101\vert x\vert011101\rangle =\langle a_{1}\vert x\vert e_{x}\rangle,\\
\langle\Phi_{A_{2};1,-1}^{c}\vert y\vert\Phi_{E,y;1,-1}^{c}\rangle & =\langle110101\vert y\vert011101\rangle =\langle a_{1}\vert y\vert e_{x}\rangle=0.
\end{align}
To summarize, the selection rules for optical transitions are
\begin{eqnarray}
\vert\Phi_{A_{2};1,0}^{c}\rangle & \underset{\rightleftharpoons}{y} & \vert\Phi_{E,x;1,0}^{c}\rangle,\\
\vert\Phi_{A_{2};1,0}^{c}\rangle & \underset{\rightleftharpoons}{x} & \vert\Phi_{E,y;1,0}^{c}\rangle,\\
\vert\Phi_{A_{2};1,1}^{c}\rangle & \underset{\rightleftharpoons}{y} & \vert\Phi_{E,x;1,1}^{c}\rangle,\\
\vert\Phi_{A_{2};1,1}^{c}\rangle & \underset{\rightleftharpoons}{x} & \vert\Phi_{E,y;1,1}^{c}\rangle,\\
\vert\Phi_{A_{2};1,-1}^{c}\rangle & \underset{\rightleftharpoons}{y} & \vert\Phi_{E,x;1,-1}^{c}\rangle,\\
\vert\Phi_{A_{2};1,-1}^{c}\rangle & \underset{\rightleftharpoons}{x} & \vert\Phi_{E,y;1,-1}^{c}\rangle,
\end{eqnarray}
where the label over the arrow indicates the polarization of the electric
field. In short,
\begin{eqnarray}
\vert\Phi_{A_{2};S,m_{s}}^{c}\rangle & \underset{\rightleftharpoons}{\alpha^{\prime}} & \vert\Phi_{E,\alpha;S,m_{s}}^{c}\rangle,
\end{eqnarray}
where $\alpha,\alpha^{\prime}=x,y$ and $\alpha\neq\alpha^{\prime}$.
In addition, $\vert\Phi_{E,x;S,m_{s}}^{c}\rangle$ and $\vert\Phi_{E,y;S,m_{s}}^{c}\rangle$
remain degenerate when applying a magnetic field.

For the electronic ground states, they can be formally diagonalized as
\begin{align}
\vert g_{i}\rangle  =\sum_{j=-1}^{1}C_{i,j}^{g}\vert\Phi_{A_{2};1,j}^{c}\rangle=C_{i,0}^{g}\vert\Phi_{A_{2};1,0}^{c}\rangle+C_{i,1}^{g}\vert\Phi_{A_{2};1,1}^{c}\rangle+C_{i,-1}^{g}\vert\Phi_{A_{2};1,-1}^{c}\rangle,
\end{align}
with eigenenergies $E_{i}^{g}$. For the electronic excited state,
they can be likewise diagonalized in two sets according to their polarizations as
\begin{align}
\vert e_{i}^{x}\rangle & =\sum_{j=-1}^{1}C_{i,j}^{e}\vert\Phi_{E,x;1,j}^{c}\rangle =C_{i,0}^{e}\vert\Phi_{E,x;1,0}^{c}\rangle+C_{i,1}^{e}\vert\Phi_{E,x;1,1}^{c}\rangle+C_{i,-1}^{e}\vert\Phi_{E,x;1,-1}^{c}\rangle,\\
\vert e_{i}^{y}\rangle & =\sum_{j=-1}^{1}C_{i,j}^{e}\vert\Phi_{E,y;1,j}^{c}\rangle =C_{i,0}^{e}\vert\Phi_{E,y;1,0}^{c}\rangle+C_{i,1}^{e}\vert\Phi_{E,y;1,1}^{c}\rangle+C_{i,-1}^{e}\vert\Phi_{E,y;1,-1}^{c}\rangle,
\end{align}
with eigenenergies $E_{i}^{e}$, where they share the same coefficients
due to the degeneracy.

According to Refs.~\cite{Jackson1999,Landau1995},
the constitutive relation reads
\begin{equation}
\vec{D}=\epsilon_{0}\overleftrightarrow{\epsilon_{r}}\vec{E}=\epsilon_{D}\epsilon_{0}\vec{E}+\vec{P},
\end{equation}
where $\epsilon_{0}$ is the permittivity of the vacuum, $\overleftrightarrow{\epsilon_{r}}$
is the relative permittivity of diamond with NV centers, $\vec{E}$
is the applied electric field, $\epsilon_{D}$ is the relative permittivity
of pure diamond, and the polarization density can be calculated by
linear response theory \cite{Kubo1985} as
\begin{eqnarray}
\vec{P} & = & -\frac{n_0}{\hbar }\mathrm{Re}\left[\sum_{j,i,f}\rho_{i}\frac{\vec{d}_{if}^{(j)}(\vec{d}_{fi}^{(j)}\cdot\vec{E})}{\omega-\Delta_{fi}^{(j)}+i\gamma}\right]\nonumber \\
 & = & -\frac{n_0}{3\hbar }\mathrm{Re}\left[\frac{\vec{d}_{g_{1}e_{1}}^{(j)}(\vec{d}_{e_{1}g_{1}}^{(j)}\cdot\vec{E})}{\omega-\Delta_{e_{1}g_{1}}^{(j)}+i\gamma}+\frac{\vec{d}_{g_{1}e_{2}}^{(j)}(\vec{d}_{e_{2}g_{1}}^{(j)}\cdot\vec{E})}{\omega-\Delta_{e_{2}g_{1}}^{(j)}+i\gamma}+\frac{\vec{d}_{g_{1}e_{3}}^{(j)}(\vec{d}_{e_{3}g_{1}}^{(j)}\cdot\vec{E})}{\omega-\Delta_{e_{3}g_{1}}^{(j)}+i\gamma}\right.\nonumber \\
 &  & +\frac{\vec{d}_{g_{2}e_{1}}^{(j)}(\vec{d}_{e_{1}g_{2}}^{(j)}\cdot\vec{E})}{\omega-\Delta_{e_{1}g_{2}}^{(j)}+i\gamma}+\frac{\vec{d}_{g_{2}e_{2}}^{(j)}(\vec{d}_{e_{2}g_{2}}^{(j)}\cdot\vec{E})}{\omega-\Delta_{e_{2}g_{2}}^{(j)}+i\gamma}+\frac{\vec{d}_{g_{2}e_{3}}^{(j)}(\vec{d}_{e_{3}g_{2}}^{(j)}\cdot\vec{E})}{\omega-\Delta_{e_{3}g_{2}}^{(j)}+i\gamma}\nonumber \\
 &  & \left.+\frac{\vec{d}_{g_{3}e_{1}}^{(j)}(\vec{d}_{e_{1}g_{3}}^{(j)}\cdot\vec{E})}{\omega-\Delta_{e_{1}g_{3}}^{(j)}+i\gamma}+\frac{\vec{d}_{g_{3}e_{2}}^{(j)}(\vec{d}_{e_{2}g_{3}}^{(j)}\cdot\vec{E})}{\omega-\Delta_{e_{2}g_{3}}^{(j)}+i\gamma}+\frac{\vec{d}_{g_{3}e_{3}}^{(j)}(\vec{d}_{e_{3}g_{3}}^{(j)}\cdot\vec{E})}{\omega-\Delta_{e_{3}g_{3}}^{(j)}+i\gamma}\right].
\end{eqnarray}
Here, $\vec{d}_{if}^{(j)}=\left\langle i\right|\vec{d}^{(j)}\left|f\right\rangle $
is the matrix element of electric dipole of $j$th NV center between
the initial state $\left|i\right\rangle $ and final state $\left|f\right\rangle $;
$\Delta_{fi}^{(j)}$ is the transition energy between the initial
state $\left|i\right\rangle $ and the final state $\left|f\right\rangle $
of $j$th NV center; and $\gamma$ is the decay rate of the electronic
excited state. The summation $\sum_{j}$ is over all NV centers within
the volume $v_{0}=n^{-1}_0$, and $\omega$ is the frequency of the incident
light. In the above equation, we did not explicitly discriminate the
contributions from $\vert e_{i}^{x}\rangle$ and $\vert e_{i}^{y}\rangle$
as they only differ by the polarization direction.

For the transition $E_{1}^{g}(j)\leftrightarrow E_{1}^{e}(j)$,
note that
\begin{align}
\frac{\vec{d}_{g_{1}e_{1}}^{(j)}(\vec{d}_{e_{1}g_{1}}^{(j)}\cdot\vec{E})}{\omega-\Delta_{e_{1}g_{1}}^{(j)}+i\gamma} & =\frac{\sum_{j_{1},j_{2},j_{3},j_{4}}C_{1,j_{1}}^{g*}C_{1,j_{2}}^{e}C_{1,j_{3}}^{e*}C_{1,j_{4}}^{g}\langle\Phi_{A_{2};1,j_{1}}^{c}\vert\vec{d}^{(j)}\vert\Phi_{E;1,j_{2}}^{c}\rangle(\langle\Phi_{E;1,j_{3}}^{c}\vert\vec{d}^{(j)}\vert\Phi_{A_{2};1,j_{4}}^{c}\rangle\cdot\vec{E})}{\omega-\left[E_{1}^{e}(j)-E_{1}^{g}(j)\right]+i\gamma}\nonumber \\
 & =\frac{\sum_{j_{1},j_{3}}C_{1,j_{1}}^{g*}C_{1,j_{1}}^{e}C_{1,j_{3}}^{e*}C_{1,j_{3}}^{g}\langle\Phi_{A_{2};1,j_{1}}^{c}\vert\vec{d}^{(j)}\vert\Phi_{E;1,j_{1}}^{c}\rangle(\langle\Phi_{E;1,j_{3}}^{c}\vert\vec{d}^{(j)}\vert\Phi_{A_{2};1,j_{3}}^{c}\rangle\cdot\vec{E})}{\omega-\left[E_{1}^{e}(j)-E_{1}^{g}(j)\right]+i\gamma}\nonumber \\
 & =\frac{\sum_{j_{1},j_{2}}C_{1,j_{1}}^{g*}C_{1,j_{1}}^{e}C_{1,j_{2}}^{e*}C_{1,j_{2}}^{g}}{\omega-\left[E_{1}^{e}(j)-E_{1}^{g}(j)\right]+i\gamma}(\vec{d}_{x}^{(j)}+\vec{d}_{y}^{(j)})(\vec{d}_{x}^{(j)}+\vec{d}_{y}^{(j)})\cdot\vec{E},\\
\frac{\vec{d}_{g_{1}e_{2}}^{(j)}(\vec{d}_{e_{2}g_{1}}^{(j)}\cdot\vec{E})}{\omega-\Delta_{e_{2}g_{1}}^{(j)}+i\gamma} & =\frac{\sum_{j_{1},j_{2}}C_{1,j_{1}}^{g*}C_{2,j_{1}}^{e}C_{2,j_{2}}^{e*}C_{1,j_{2}}^{g}}{\omega-\left[E_{2}^{e}(j)-E_{1}^{g}(j)\right]+i\gamma}(\vec{d}_{x}^{(j)}+\vec{d}_{y}^{(j)})(\vec{d}_{x}^{(j)}+\vec{d}_{y}^{(j)})\cdot\vec{E},\\
\frac{\vec{d}_{g_{1}e_{3}}^{(j)}(\vec{d}_{e_{3}g_{1}}^{(j)}\cdot\vec{E})}{\omega-\Delta_{e_{3}g_{1}}^{(j)}+i\gamma} & =\frac{\sum_{j_{1},j_{2}}C_{1,j_{1}}^{g*}C_{3,j_{1}}^{e}C_{3,j_{2}}^{e*}C_{1,j_{2}}^{g}}{\omega-\left[E_{3}^{e}(j)-E_{1}^{g}(j)\right]+i\gamma}(\vec{d}_{x}^{(j)}+\vec{d}_{y}^{(j)})(\vec{d}_{x}^{(j)}+\vec{d}_{y}^{(j)})\cdot\vec{E},
\end{align}
\begin{align}
\frac{\vec{d}_{g_{2}e_{1}}^{(j)}(\vec{d}_{e_{1}g_{2}}^{(j)}\cdot\vec{E})}{\omega-\Delta_{e_{1}g_{2}}^{(j)}+i\gamma} & =\frac{\sum_{j_{1},j_{2}}C_{2,j_{1}}^{g*}C_{1,j_{1}}^{e}C_{1,j_{2}}^{e*}C_{2,j_{2}}^{g}}{\omega-\left[E_{1}^{e}(j)-E_{2}^{g}(j)\right]+i\gamma}(\vec{d}_{x}^{(j)}+\vec{d}_{y}^{(j)})(\vec{d}_{x}^{(j)}+\vec{d}_{y}^{(j)})\cdot\vec{E},\\
\frac{\vec{d}_{g_{2}e_{2}}^{(j)}(\vec{d}_{e_{2}g_{2}}^{(j)}\cdot\vec{E})}{\omega-\Delta_{e_{2}g_{2}}^{(j)}+i\gamma} & =\frac{\sum_{j_{1},j_{2}}C_{2,j_{1}}^{g*}C_{2,j_{1}}^{e}C_{2,j_{2}}^{e*}C_{2,j_{2}}^{g}}{\omega-\left[E_{2}^{e}(j)-E_{2}^{g}(j)\right]+i\gamma}(\vec{d}_{x}^{(j)}+\vec{d}_{y}^{(j)})(\vec{d}_{x}^{(j)}+\vec{d}_{y}^{(j)})\cdot\vec{E},\\
\frac{\vec{d}_{g_{2}e_{3}}^{(j)}(\vec{d}_{e_{3}g_{2}}^{(j)}\cdot\vec{E})}{\omega-\Delta_{e_{3}g_{2}}^{(j)}+i\gamma} & =\frac{\sum_{j_{1},j_{2}}C_{2,j_{1}}^{g*}C_{3,j_{1}}^{e}C_{3,j_{2}}^{e*}C_{2,j_{2}}^{g}}{\omega-\left[E_{3}^{e}(j)-E_{2}^{g}(j)\right]+i\gamma}(\vec{d}_{x}^{(j)}+\vec{d}_{y}^{(j)})(\vec{d}_{x}^{(j)}+\vec{d}_{y}^{(j)})\cdot\vec{E},
\end{align}
\begin{align}
\frac{\vec{d}_{g_{3}e_{1}}^{(j)}(\vec{d}_{e_{1}g_{3}}^{(j)}\cdot\vec{E})}{\omega-\Delta_{e_{1}g_{3}}^{(j)}+i\gamma} & =\frac{\sum_{j_{1},j_{2}}C_{3,j_{1}}^{g*}C_{1,j_{1}}^{e}C_{1,j_{2}}^{e*}C_{3,j_{2}}^{g}}{\omega-\left[E_{1}^{e}(j)-E_{3}^{g}(j)\right]+i\gamma}(\vec{d}_{x}^{(j)}+\vec{d}_{y}^{(j)})(\vec{d}_{x}^{(j)}+\vec{d}_{y}^{(j)})\cdot\vec{E},\\
\frac{\vec{d}_{g_{3}e_{2}}^{(j)}(\vec{d}_{e_{2}g_{3}}^{(j)}\cdot\vec{E})}{\omega-\Delta_{e_{2}g_{3}}^{(j)}+i\gamma} & =\frac{\sum_{j_{1},j_{2}}C_{3,j_{1}}^{g*}C_{2,j_{1}}^{e}C_{2,j_{2}}^{e*}C_{3,j_{2}}^{g}}{\omega-\left[E_{2}^{e}(j)-E_{3}^{g}(j)\right]+i\gamma}(\vec{d}_{x}^{(j)}+\vec{d}_{y}^{(j)})(\vec{d}_{x}^{(j)}+\vec{d}_{y}^{(j)})\cdot\vec{E},\\
\frac{\vec{d}_{g_{3}e_{3}}^{(j)}(\vec{d}_{e_{3}g_{3}}^{(j)}\cdot\vec{E})}{\omega-\Delta_{e_{3}g_{3}}^{(j)}+i\gamma} & =\frac{\sum_{j_{1},j_{2}}C_{3,j_{1}}^{g*}C_{3,j_{1}}^{e}C_{3,j_{2}}^{e*}C_{3,j_{2}}^{g}}{\omega-\left[E_{3}^{e}(j)-E_{3}^{g}(j)\right]+i\gamma}(\vec{d}_{x}^{(j)}+\vec{d}_{y}^{(j)})(\vec{d}_{x}^{(j)}+\vec{d}_{y}^{(j)})\cdot\vec{E}.
\end{align}
Therefore, the induced dielectric polarization density can be written as
\begin{eqnarray}
\vec{P} & = & -\frac{n_{0}}{3\hbar}\mathrm{Re}\left\{ \frac{\sum_{j_{1},j_{2}}C_{1,j_{1}}^{g*}C_{1,j_{1}}^{e}C_{1,j_{2}}^{e*}C_{1,j_{2}}^{g}}{\omega-\left[E_{1}^{e}(j)-E_{1}^{g}(j)\right]+i\gamma}+\frac{\sum_{j_{1},j_{2}}C_{1,j_{1}}^{g*}C_{2,j_{1}}^{e}C_{2,j_{2}}^{e*}C_{1,j_{2}}^{g}}{\omega-\left[E_{2}^{e}(j)-E_{1}^{g}(j)\right]+i\gamma}+\frac{\sum_{j_{1},j_{2}}C_{1,j_{1}}^{g*}C_{3,j_{1}}^{e}C_{3,j_{2}}^{e*}C_{1,j_{2}}^{g}}{\omega-\left[E_{3}^{e}(j)-E_{1}^{g}(j)\right]+i\gamma}\right.\nonumber \\
 &  & +\frac{\sum_{j_{1},j_{2}}C_{2,j_{1}}^{g*}C_{1,j_{1}}^{e}C_{1,j_{2}}^{e*}C_{2,j_{2}}^{g}}{\omega-\left[E_{1}^{e}(j)-E_{2}^{g}(j)\right]+i\gamma}+\frac{\sum_{j_{1},j_{2}}C_{2,j_{1}}^{g*}C_{2,j_{1}}^{e}C_{2,j_{2}}^{e*}C_{2,j_{2}}^{g}}{\omega-\left[E_{2}^{e}(j)-E_{2}^{g}(j)\right]+i\gamma}+\frac{\sum_{j_{1},j_{2}}C_{2,j_{1}}^{g*}C_{3,j_{1}}^{e}C_{3,j_{2}}^{e*}C_{2,j_{2}}^{g}}{\omega-\left[E_{3}^{e}(j)-E_{2}^{g}(j)\right]+i\gamma}\nonumber \\
 &  & \left.+\frac{\sum_{j_{1},j_{2}}C_{3,j_{1}}^{g*}C_{1,j_{1}}^{e}C_{1,j_{2}}^{e*}C_{3,j_{2}}^{g}}{\omega-\left[E_{1}^{e}(j)-E_{3}^{g}(j)\right]+i\gamma}+\frac{\sum_{j_{1},j_{2}}C_{3,j_{1}}^{g*}C_{2,j_{1}}^{e}C_{2,j_{2}}^{e*}C_{3,j_{2}}^{g}}{\omega-\left[E_{2}^{e}(j)-E_{3}^{g}(j)\right]+i\gamma}+\frac{\sum_{j_{1},j_{2}}C_{3,j_{1}}^{g*}C_{3,j_{1}}^{e}C_{3,j_{2}}^{e*}C_{3,j_{2}}^{g}}{\omega-\left[E_{3}^{e}(j)-E_{3}^{g}(j)\right]+i\gamma}\right\} \nonumber \\
 &  & \times(\vec{d}_{x}^{(j)}+\vec{d}_{y}^{(j)})(\vec{d}_{x}^{(j)}+\vec{d}_{y}^{(j)})\cdot\vec{E}.
\end{eqnarray}
As a result, the relative permittivity tensor is
\begin{eqnarray}
\overleftrightarrow{\epsilon_{r}}(\omega) & = & \epsilon_{D}-\frac{n_{0}}{3\hbar\epsilon_{0}}\mathrm{Re}\left\{ \frac{\sum_{j_{1},j_{2}}C_{1,j_{1}}^{g*}C_{1,j_{1}}^{e}C_{1,j_{2}}^{e*}C_{1,j_{2}}^{g}}{\omega-\left[E_{1}^{e}(j)-E_{1}^{g}(j)\right]+i\gamma}+\frac{\sum_{j_{1},j_{2}}C_{1,j_{1}}^{g*}C_{2,j_{1}}^{e}C_{2,j_{2}}^{e*}C_{1,j_{2}}^{g}}{\omega-\left[E_{2}^{e}(j)-E_{1}^{g}(j)\right]+i\gamma}+\frac{\sum_{j_{1},j_{2}}C_{1,j_{1}}^{g*}C_{3,j_{1}}^{e}C_{3,j_{2}}^{e*}C_{1,j_{2}}^{g}}{\omega-\left[E_{3}^{e}(j)-E_{1}^{g}(j)\right]+i\gamma}\right.\nonumber \\
 &  & +\frac{\sum_{j_{1},j_{2}}C_{2,j_{1}}^{g*}C_{1,j_{1}}^{e}C_{1,j_{2}}^{e*}C_{2,j_{2}}^{g}}{\omega-\left[E_{1}^{e}(j)-E_{2}^{g}(j)\right]+i\gamma}+\frac{\sum_{j_{1},j_{2}}C_{2,j_{1}}^{g*}C_{2,j_{1}}^{e}C_{2,j_{2}}^{e*}C_{2,j_{2}}^{g}}{\omega-\left[E_{2}^{e}(j)-E_{2}^{g}(j)\right]+i\gamma}+\frac{\sum_{j_{1},j_{2}}C_{2,j_{1}}^{g*}C_{3,j_{1}}^{e}C_{3,j_{2}}^{e*}C_{2,j_{2}}^{g}}{\omega-\left[E_{3}^{e}(j)-E_{2}^{g}(j)\right]+i\gamma}\nonumber \\
 &  & \left.+\frac{\sum_{j_{1},j_{2}}C_{3,j_{1}}^{g*}C_{1,j_{1}}^{e}C_{1,j_{2}}^{e*}C_{3,j_{2}}^{g}}{\omega-\left[E_{1}^{e}(j)-E_{3}^{g}(j)\right]+i\gamma}+\frac{\sum_{j_{1},j_{2}}C_{3,j_{1}}^{g*}C_{2,j_{1}}^{e}C_{2,j_{2}}^{e*}C_{3,j_{2}}^{g}}{\omega-\left[E_{2}^{e}(j)-E_{3}^{g}(j)\right]+i\gamma}+\frac{\sum_{j_{1},j_{2}}C_{3,j_{1}}^{g*}C_{3,j_{1}}^{e}C_{3,j_{2}}^{e*}C_{3,j_{2}}^{g}}{\omega-\left[E_{3}^{e}(j)-E_{3}^{g}(j)\right]+i\gamma}\right\} \nonumber \\
 &  & \times(\vec{d}_{x}^{(j)}+\vec{d}_{y}^{(j)})(\vec{d}_{x}^{(j)}+\vec{d}_{y}^{(j)}).
 \label{eq:Epsilon}
\end{eqnarray}
Clearly, there may be nine possible negative permittivities around
the nine transition frequencies $\Delta_{e_{i}g_{j}}^{(j)}=E_{i}^{e}(j)-E_{j}^{g}(j)$.

\begin{figure}
\includegraphics[bb=100bp 300bp 315bp 480bp,width=6cm,angle=0]{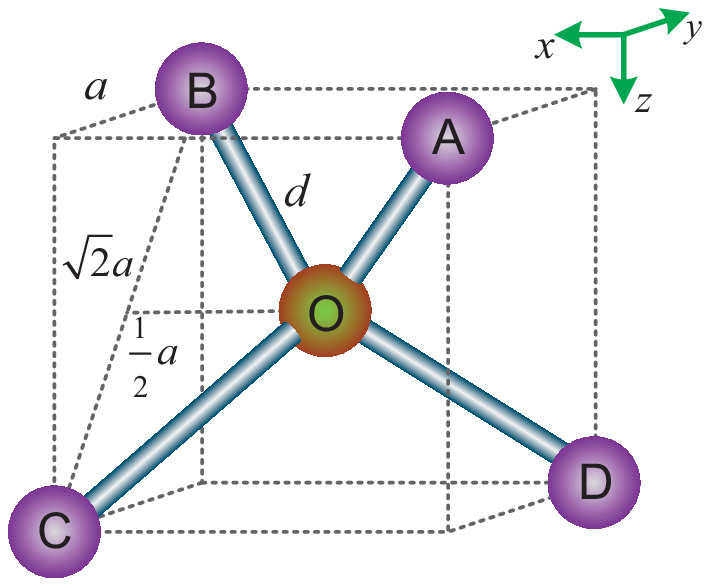}
\caption{(color online) Four possible orientations of NV centers in diamond
\cite{Zou2014,Doherty2013}: $\vec{r}_\mathrm{OA}=(-1,-1,-1)/\sqrt{3}$, $\vec{r}_\mathrm{OB}=(1,1,-1)/\sqrt{3}$,
$\vec{r}_\mathrm{OC}=(1,-1,1)/\sqrt{3}$, and $\vec{r}_\mathrm{OD}=(-1,1,1)/\sqrt{3}$. $d=154$
pm is the length of carbon bond. The side length of the cube is $a=\frac{2}{\sqrt{3}}d$.
The angle between any pair of the
above four orientations is $109\protect\textdegree28'$.\label{fig:Orientation}}
\end{figure}

As shown in Fig.~\ref{fig:Orientation}, in diamond there are four
possible symmetry axes for the NV centers, i.e. $\vec{r}_\textrm{OA}=(-1,-1,-1)/\sqrt{3}$,
$\vec{r}_\textrm{OB}=(1,1,-1)/\sqrt{3}$, $\vec{r}_\textrm{OC}=(1,-1,1)/\sqrt{3}$, and
$\vec{r}_\textrm{OD}=(-1,1,1)/\sqrt{3}$.
Here, $\vec{r}_\textrm{OA}$ can be obtained by rotating the $z$-axis around the axis
$\vec{n}_\textrm{OA}=(-1,1,0)/\sqrt{2}$ by an angle $\theta_\textrm{OA}=-(180\textdegree-109\textdegree28'/2)=-125\textdegree16'$,
i.e.,
\begin{equation}
\vec{r}_\textrm{OA}=R(\vec{n}_\textrm{OA},\theta_\textrm{OA})\hat{e}_{z}=R(\vec{n}_\textrm{OA},\theta_\textrm{OA})(0,0,1)^{T},
\end{equation}
where the rotation matrix around $\vec{n}=(n_{x},n_{y},n_{z})$ with
an angle $\theta$ is \cite{Sakurai1993}
\begin{equation}
R(\vec{n},\theta)=\begin{pmatrix}\cos\theta+n_{x}^{2}(1-\cos\theta) & n_{x}n_{y}(1-\cos\theta)-n_{z}\sin\theta & n_{x}n_{z}(1-\cos\theta)+n_{y}\sin\theta\\
n_{x}n_{y}(1-\cos\theta)+n_{z}\sin\theta & \cos\theta+n_{y}^{2}(1-\cos\theta) & n_{y}n_{z}(1-\cos\theta)-n_{x}\sin\theta\\
n_{x}n_{z}(1-\cos\theta)-n_{y}\sin\theta & n_{y}n_{z}(1-\cos\theta)+n_{x}\sin\theta & \cos\theta+n_{z}^{2}(1-\cos\theta)
\end{pmatrix}.
\end{equation}
And $\vec{r}_\textrm{OB}$ can be obtained by rotating the $z$-axis around the axis
$\vec{n}_\textrm{OA}=(-1,1,0)/\sqrt{2}$ by an angle $-\theta_\textrm{OA}=125\textdegree16'$,
i.e.,
\begin{equation}
\vec{r}_\textrm{OB}=R(\vec{n}_\textrm{OA},-\theta_\textrm{OA})\hat{e}_{z}.
\end{equation}
And $\vec{r}_\textrm{OC}$ can be obtained by rotating the $z$-axis around the axis
$\vec{n}_\textrm{OC}=(1,1,0)/\sqrt{2}$ by an angle $\theta_\textrm{OC}=109\textdegree28'/2=54\textdegree44'$,
i.e.,
\begin{equation}
\vec{r}_\textrm{OC}=R(\vec{n}_\textrm{OC},\theta_\textrm{OC})\hat{e}_{z}.
\end{equation}
And $\vec{r}_\textrm{OD}$ can be obtained by rotating the $z$-axis around the axis
$\vec{n}_\textrm{OC}=(1,1,0)/\sqrt{2}$ by an angle $-\theta_\textrm{OC}=-54\textdegree44'$,
i.e.,
\begin{equation}
\vec{r}_\textrm{OD}=R(\vec{n}_\textrm{OC},-\theta_\textrm{OC})\hat{e}_{z}.
\end{equation}

\section{Quantum Switch of Negative Refraction and Normal Refraction}
\label{sec:Switch}

\begin{figure}
\includegraphics[width=9cm]{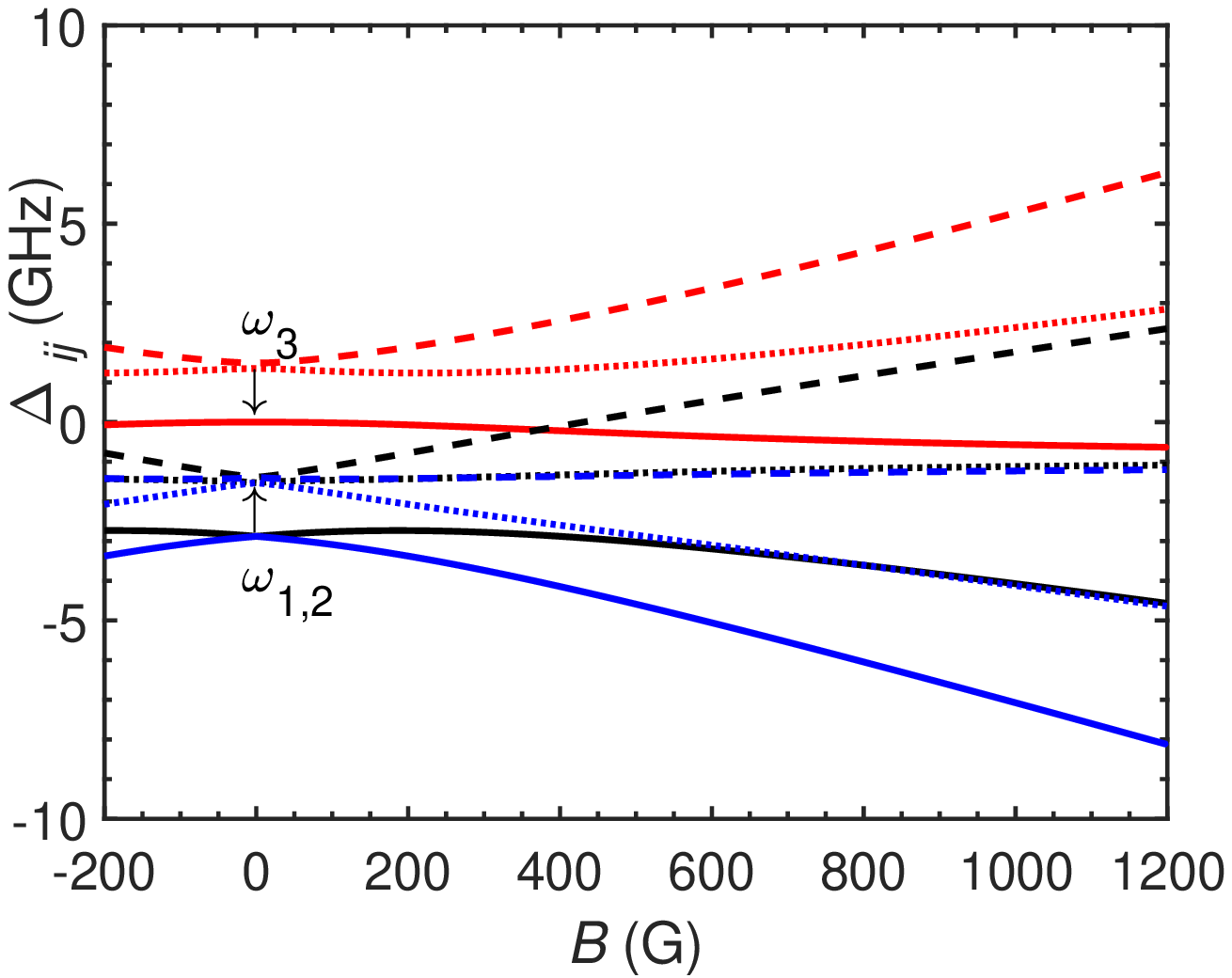}
\caption{The transition frequencies $\Delta_{ij}=E^e_i-E^g_j$
versus the magnetic field $B$. The magnetic field is applied along
the $z$-axis. This figure is identical for all four possible orientations
of the NV centers due to the special choice of $\vec{B}\parallel\vec{e}_z$.\label{fig:TranFreqVsB_n0}}
\end{figure}

As implied by Eq.~(\ref{eq:Epsilon}), the permittivity might be negative around the transition frequencies. In order to switch on/off the negative refraction,
we analyze the effect of the magnetic field on the transition frequencies.
As shown in Fig.~\ref{fig:TranFreqVsB_n0},
there are generally nine transition frequencies $\Delta_{ij}=E^e_{i}-E^g_{j}$,
which can be subtly tuned by varying the magnetic field.

\begin{figure}
\includegraphics[width=8.5cm]{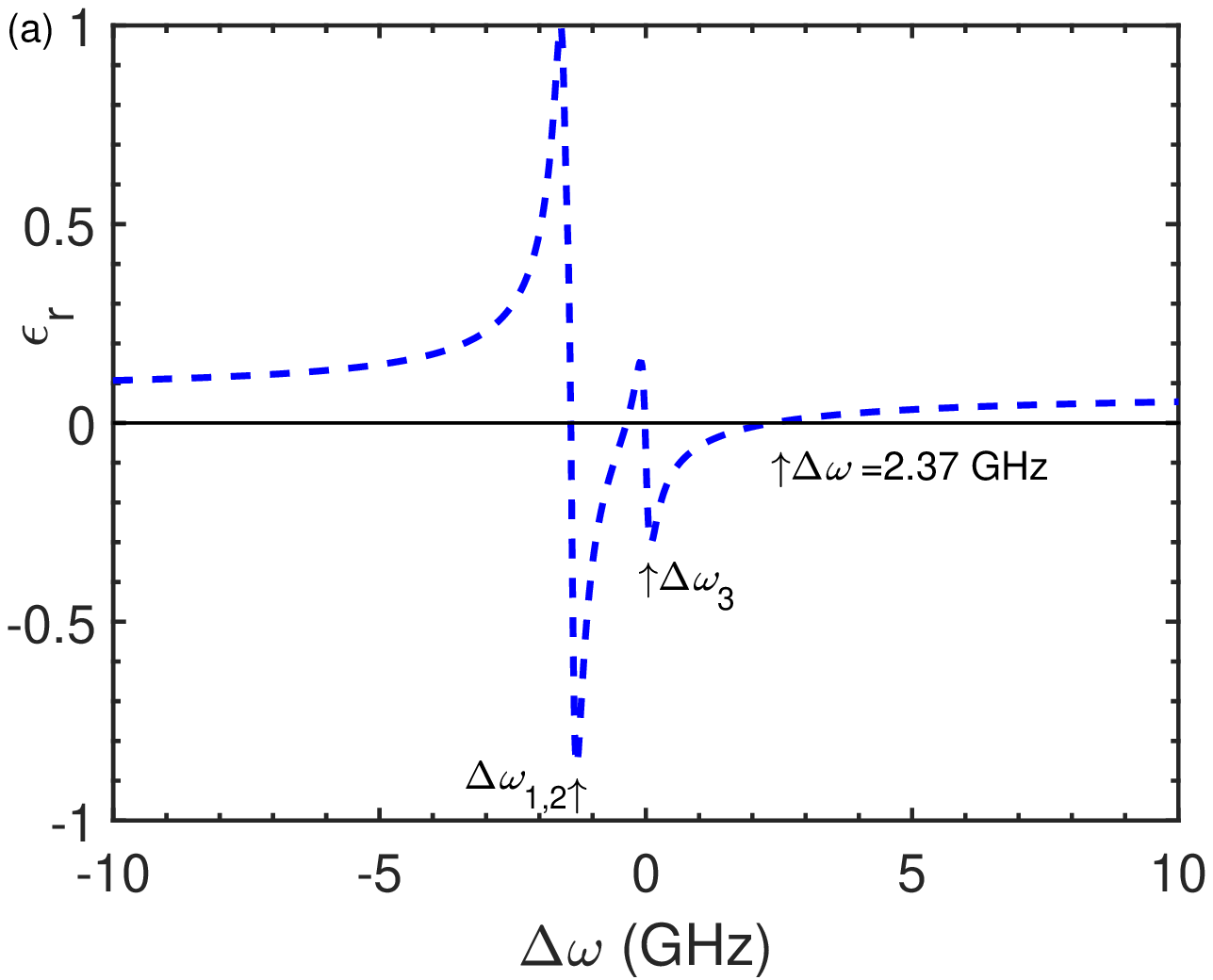}
\includegraphics[width=8.5cm]{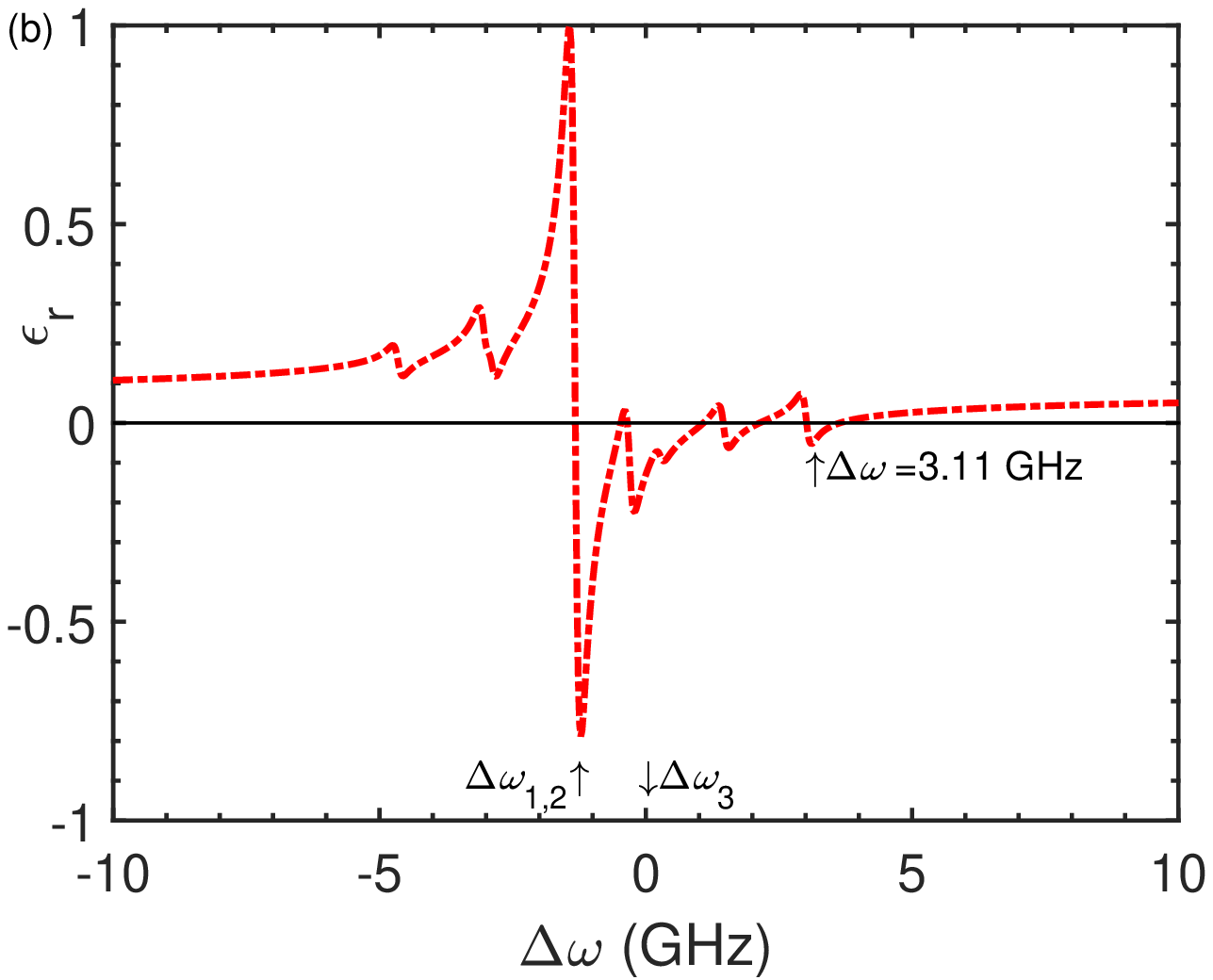}
\includegraphics[width=8.5cm]{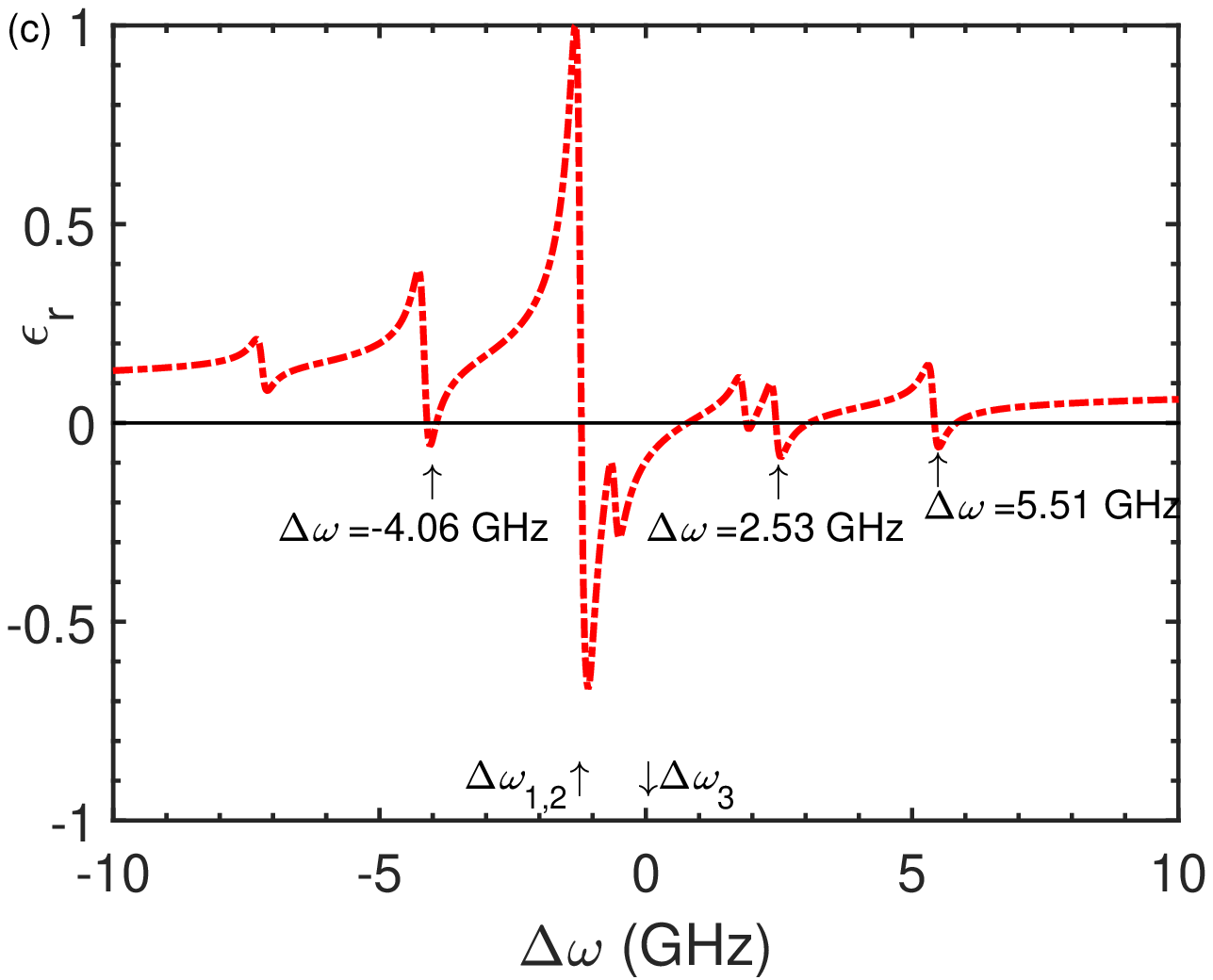}
\includegraphics[width=8.5cm]{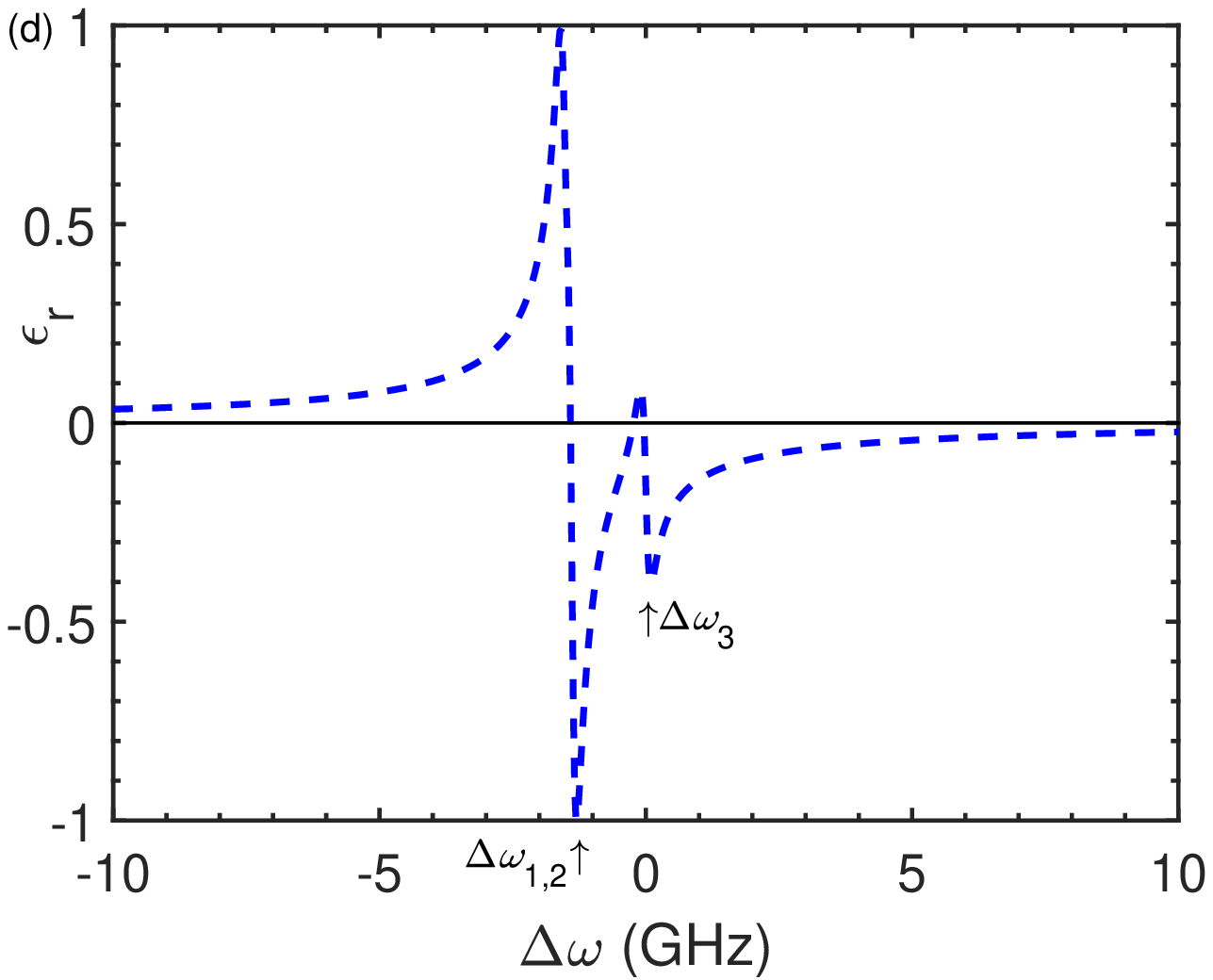}
\caption{(color online) The frequency dependence of the electric permittivity $\protect\overleftrightarrow{\epsilon_{r}}$
of diamond with NV centers for different values of the magnetic field $B$
and density of NV centers $n_{0}$: (a) $B_z=0$~G and $n_{0}=0.5$~ppm, (b) $B_z=514$~G and $n_{0}=0.5$~ppm, (c) $B_z=1025$~G and $n_{0}=0.5$~ppm, (d) $B_z=0$~G and $n_{0}=16$~ppm. Other
parameters are $d_{x}=d_{y}=11$~D \cite{Lenef1996}, $\gamma^{-1}=10$
ns \cite{Acosta2011}, $\epsilon_{D}=5.7$ \cite{Fontanella1977},
and $\mu_{D}=1-2.1\times10^{-5}$ \cite{Young1992}, $B_x=B_y=0$ G.
The thin black line $\protect\epsilon_{r}=0$ is just a guide to the eye.\label{fig:EpVsB_n0}}
\end{figure}

Below, we analyze the NV center at some specific magnetic fields:

(a) When the magnetic field is absent, the Hamiltonians of the electronic
ground state and excited state are respectively simplified as
\begin{eqnarray}
H_{\mathrm{gs}} & = & D_{\mathrm{gs}}\sum_{m_{s}=\pm1}\vert\Phi_{A_{2};1,m_{s}}^{c}\rangle\langle\Phi_{A_{2};1,m_{s}}^{c}\vert,\\
H_{\mathrm{es}} & = & \sum_{\alpha=x,y}[D_{\mathrm{es}}^{\parallel}(\vert\Phi_{E,\alpha;1,+1}^{c}\rangle\langle\Phi_{E,\alpha;1,+1}^{c}\vert+\vert\Phi_{E,\alpha;1,-1}^{c}\rangle\langle\Phi_{E,\alpha;1,-1}^{c}\vert)-\xi\vert\Phi_{E,\alpha;1,-1}^{c}\rangle\langle\Phi_{E,\alpha;1,-1}^{c}\vert+\mathrm{h.c.}].
\end{eqnarray}
The eigenstates of the electronic ground state are
\begin{align}
\vert g_{1}\rangle & =\vert\Phi_{A_{2};1,0}^{c}\rangle,\\
\vert g_{2}\rangle & =\vert\Phi_{A_{2};1,+1}^{c}\rangle,\\
\vert g_{3}\rangle & =\vert\Phi_{A_{2};1,-1}^{c}\rangle,
\end{align}
with eigenenergies
\begin{align}
E^g_{1} & =0,\\
E^g_{2} & =D_{\mathrm{gs}},\\
E^g_{3} & =D_{\mathrm{gs}}.
\end{align}
The eigenstates of the electronic excited state are
\begin{align}
\vert e_{1}^{\alpha}\rangle & =\vert\Phi_{E,\alpha;S,0}^{c}\rangle,\\
\vert e_{2}^{\alpha}\rangle & =(\vert\Phi_{E,\alpha;S,+1}^{c}\rangle+\vert\Phi_{E,\alpha;S,-1}^{c}\rangle)/\sqrt{2},\\
\vert e_{3}^{\alpha}\rangle & =(\vert\Phi_{E,\alpha;S,+1}^{c}\rangle-\vert\Phi_{E,\alpha;S,-1}^{c}\rangle)/\sqrt{2},
\end{align}
with eigenenergies
\begin{align}
E^e_{1} & =0,\\
E^e_{2} & =D_{\mathrm{es}}^{\parallel}+\xi,\\
E^e_{3} & =D_{\mathrm{es}}^{\parallel}-\xi.
\end{align}
Because the spin is conserved, the following optical transitions are
allowed
\begin{align}
\vert g_{1}\rangle & \rightleftharpoons\vert e_{1}\rangle,\\
\vert g_{2}\rangle,\vert g_{3}\rangle & \rightleftharpoons\vert e_{2}\rangle,\vert e_{3}\rangle.
\end{align}
These correspond to five peaks around the transition frequencies $\Delta_{11}$,
$\Delta_{22}$, $\Delta_{23}$, $\Delta_{32}$,
and $\Delta_{33}$. Furthermore, because $\vert g_{2}\rangle$
and $\vert g_{3}\rangle$ are degenerate, $\Delta_{22}=\Delta_{23}=D_{\mathrm{es}}^{\parallel}-D_{\mathrm{gs}}+\xi$
and $\Delta_{32}=\Delta_{33}=D_{\mathrm{es}}^{\parallel}-D_{\mathrm{gs}}-\xi$.
Since the separations between the latter four peaks are $2\xi$,
which is, of the order of GHz, much smaller than the width of the peaks.
In Fig.~\ref{fig:EpVsB_n0}, we plot the permittivity for different values
of the magnetic field $B$ and density $n_{0}$ of NV centers.
Therefore, we would only observe two dips for $B=0$, as shown in Fig.~\ref{fig:EpVsB_n0}(a).
Moreover, when the frequency is larger than $\Delta\omega=2.37$~GHz,
the permittivity becomes positive and thus negative refraction disappears.
When the density is increased to $n_0=16$~ppm, cf. Fig.~\ref{fig:EpVsB_n0}(d),
the two negative dips remains but the windows on the right is significantly broadened.

(b) When $B=514$~G, the electronic excited state is at the avoided
crossing point. Correspondingly, the Hamiltonians for the electronic ground and excited states are, respectively, given by
\begin{eqnarray}
H_{\mathrm{es}}&\simeq&
\begin{pmatrix}
2D_{\mathrm{es}}^{\parallel} & D_{\mathrm{es}}^{\parallel}e^{-i\phi} & 0\\
D_{\mathrm{es}}^{\parallel}e^{i\phi} & 0 & D_{\mathrm{es}}^{\parallel}e^{-i\phi}\\
0 & D_{\mathrm{es}}^{\parallel}e^{i\phi} & 0
\end{pmatrix},\\
H_{\mathrm{gs}} & \simeq &
\begin{pmatrix}
\frac{3}{2}D_{\mathrm{gs}} & \frac{1}{2}D_{\mathrm{gs}}e^{-i\phi} & 0\\
\frac{1}{2}D_{\mathrm{gs}}e^{i\phi} & 0 & \frac{1}{2}D_{\mathrm{gs}}e^{-i\phi}\\
0 & \frac{1}{2}D_{\mathrm{gs}}e^{i\phi} & \frac{1}{2}D_{\mathrm{gs}}
\end{pmatrix}.
\end{eqnarray}
Since the couplings are comparable to the detunings,
the eigenstates in both manifolds effectively mix all three components, i.e.,
$\vert g_{j}\rangle =\sum_{m_z}a_{m_z}\vert\Phi_{A_{2};1,m_z}^{c}\rangle$,
$\vert e_{3}^{\alpha}\rangle =\sum_{m_z}b_{m_z}\vert\Phi_{E,\alpha;S,m_z}^{c}\rangle$.
In this case, we would expect nine possible negative dips at nine transition frequencies.
In Fig.~\ref{fig:EpVsB_n0}(b), we observe 7 dips
because the degeneracy has been partially broken
and there are still two sets of degenerate states.
As compared to Fig.~\ref{fig:EpVsB_n0}(a),
there is an additional negative dip at $\Delta\omega=3.11$~GHz.

(c) When $B=1025$~G, there are three additional negative dips at $\Delta\omega=-4.06$~GHz,
$\Delta\omega=2.53$~GHz, and $\Delta\omega=5.51$~GHz beyond the domain $[-1.46,2.37]$~GHz.

\section{Negative Refraction at Interface}
\label{sec:NegRefraction}

\begin{figure}
\includegraphics[bb=100bp 285bp 295bp 480bp,width=6cm]{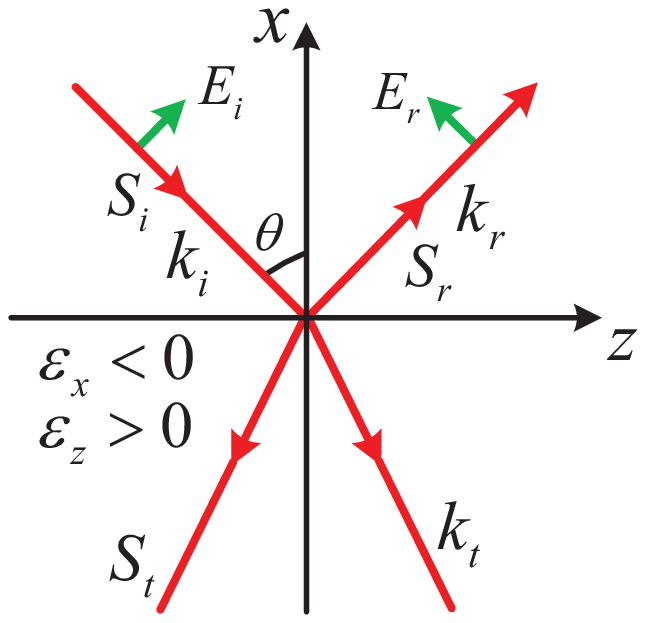}
\caption{(color online) Negative refraction for hyperbolic dispersion with
$\epsilon_{x}<0$ and $\epsilon_{z}>0$. The TH mode is incident on
the interface with electric field $\vec{E}_{i}$, wavevector $\vec{k}_{i}$,
Poynting vector $\vec{S}_{i}$, and angle $\theta$. It is reflected
with electric field $\vec{E}_{r}$, wavevector $\vec{k}_{r}$, and
Poynting vector $\vec{S}_{r}$. The Poynting vector, wavevector, electric
and magnetic fields of the transmitted wave are respectively $\vec{S}_{t}$,
$\vec{k}_{t}$, $\vec{E}_{t}$, and $\vec{H}_{t}$.\label{fig:interface}}
\end{figure}

By numerical simulation, we have shown that one principle component of the permittivity can be negative while the other principle components remain unchanged.
In this section, we will analytically prove that negative refraction can occur at the interface for a transverse magnetic (TH) mode, as shown in Fig.~\ref{fig:interface}.
According to Maxwell's equation \cite{Jackson1999,Landau1995},
\begin{align}
\nabla\times\vec{E} & =-\frac{\partial}{\partial t}\vec{B}=-\frac{\partial}{\partial t}\mu_{0}\vec{H},\\
\nabla\times\vec{H} & =\frac{\partial}{\partial t}\vec{D}=\frac{\partial}{\partial t}\overleftrightarrow{\epsilon}\vec{E},\\
\nabla\cdot\vec{D} & =0,\\
\nabla\cdot\vec{B} & =0,
\end{align}
where we have assumed $\vec{J}=0$ and $\rho=0$, $\overleftrightarrow{\epsilon}$ is
the permittivity of diamond with NV centers, $\mu_{0}$ is the permeability
of vacuum and pure diamond.

Assuming that the electric and magnetic fields of the transmitted
wave are respectively
\begin{align}
\vec{E}_{t}(\vec{r},t) & =(E_{tx}\hat{e}_{x}+E_{tz}\hat{e}_{z})\exp\left[i(\vec{k}_{t}\cdot\vec{r}-\omega t)\right],\\
\vec{H}_{t}(\vec{r},t) & =H_{ty}\hat{e}_{y}\exp\left[i(\vec{k}_{t}\cdot\vec{r}-\omega t)\right],
\end{align}
where $\vec{k}_{t}$ and $\omega$ are respectively the wavevector
and frequency of the transmitted wave, we have
\begin{align}
\nabla\times\vec{E} & =i\omega\mu_{0}\vec{H},\label{eq:E}\\
\nabla\times\vec{H} & =-i\omega\overleftrightarrow{\epsilon}\vec{E}.\label{eq:H}
\end{align}
By inserting Eq.~(\ref{eq:H}) into Eq.~(\ref{eq:E}), we have
\begin{align}
\nabla\times\nabla\times\vec{E} & =i\omega\mu_{0}\nabla\times\vec{H}\nonumber \\
 & =\overleftrightarrow{\epsilon}\mu_{0}\omega^{2}\vec{E}.
\end{align}
This is equivalent to
\begin{equation}
(\nabla\times\nabla\times\overleftrightarrow{I}-\overleftrightarrow{\epsilon}\mu_{0}\omega^{2})\vec{E}=0,\label{eq:Eigen}
\end{equation}
where $\overleftrightarrow{I}$ is the identity dyadic. For nontrivial
solutions, the determinant of the dyadic should be zero, i.e.
\begin{align}
\det\left(\begin{array}{ccc}
\mu_{0}\epsilon_{x}\omega^{2}-k_{y}^{2}-k_{z}^{2} & k_{x}k_{y} & k_{x}k_{z}\\
k_{x}k_{y} & \mu_{0}\epsilon_{y}\omega^{2}-k_{x}^{2}-k_{z}^{2} & k_{y}k_{z}\\
k_{x}k_{z} & k_{y}k_{z} & \mu_{0}\epsilon_{z}\omega^{2}-k_{x}^{2}-k_{y}^{2}
\end{array}\right) & =0,
\end{align}
or equivalently
\begin{equation}
\mu_{0}\omega^{2}\left\{ k_{x}^{2}\left[k_{y}^{2}\left(\epsilon_{x}+\epsilon_{y}\right)+k_{z}^{2}\left(\epsilon_{x}+\epsilon_{z}\right)-\mu\omega^{2}\epsilon_{x}\left(\epsilon_{y}+\epsilon_{z}\right)\right]+\left[\epsilon_{z}\left(k_{z}^{2}-\mu\omega^{2}\epsilon_{y}\right)+k_{y}^{2}\epsilon_{y}\right]\left(k_{y}^{2}+k_{z}^{2}-\mu\omega^{2}\epsilon_{x}\right)+k_{x}^{4}\epsilon_{x}\right\} =0,
\end{equation}
where
\begin{equation}
\overleftrightarrow{\epsilon}=\begin{pmatrix}\epsilon_{x} & 0 & 0\\
0 & \epsilon_{y} & 0\\
0 & 0 & \epsilon_{z}
\end{pmatrix}.
\end{equation}
The dispersion relations for the ordinary and extraordinary modes
are respectively
\begin{align}
k_{x}^{2}+k_{z}^{2}-\mu_{0}\omega^{2}\epsilon_{y} & =0,\\
\epsilon_{x}k_{x}^{2}+\epsilon_{z}k_{z}^{2}-\mu_{0}\omega^{2}\epsilon_{x}\epsilon_{z} & =0,\label{eq:ExtraOrDispersion}
\end{align}
where we have assumed $k_{y}=0$.

According to the boundary condition \cite{Belov2003}, the tangential
components of the wavevector across the interface should be equal,
i.e.
\begin{align}
k_{tz} & =k_{iz}>0,\\
k_{ty} & =k_{iy}.
\end{align}
By inserting Eq.~(\ref{eq:ExtraOrDispersion}) into Eq.~(\ref{eq:Eigen}),
we obtain the relation between $E_{tx}$ and $E_{tz}$ as
\begin{equation}
\epsilon_{x}k_{tx}E_{tx}+\epsilon_{z}k_{tz}E_{tz}=0.
\end{equation}
Using Eq.~(\ref{eq:E}), we have
\begin{align}
\vec{H} & =\frac{\nabla\times\vec{E}_{t}}{i\omega\mu_{0}}\nonumber \\
 & =\frac{\vec{k}_{t}\times\vec{E}_{t}}{\omega\mu_{0}}\nonumber \\
 & =\frac{k_{tz}E_{tx}-k_{tx}E_{tz}}{\omega\mu_{0}}\hat{e}_{y}e^{i(\vec{k}_{t}\cdot\vec{r}-\omega t)}\nonumber \\
 & =\frac{1}{\omega\mu_{0}}\left(-k_{tz}\frac{\epsilon_{z}k_{tz}E_{tz}}{\epsilon_{x}k_{tx}}-k_{tx}E_{tz}\right)\hat{e}_{y}\;e^{i(\vec{k}_{t}\cdot\vec{r}-\omega t)}\nonumber \\
 & =\frac{E_{tz}}{\omega\mu_{0}}\left(-\frac{\epsilon_{z}k_{tz}^{2}+\epsilon_{x}k_{tx}^{2}}{\epsilon_{x}k_{tx}}\right)\hat{e}_{y}\;e^{i(\vec{k}_{t}\cdot\vec{r}-\omega t)}\nonumber \\
 & =-\frac{E_{tz}}{\omega\mu_{0}}\frac{\mu_{0}\omega^{2}\epsilon_{x}\epsilon_{z}}{\epsilon_{x}k_{tx}}\;\hat{e}_{y}\;e^{i(\vec{k}_{t}\cdot\vec{r}-\omega t)}\nonumber \\
 & =-\frac{\omega\epsilon_{z}E_{tz}}{k_{tx}}\;\hat{e}_{y}\;e^{i(\vec{k}_{t}\cdot\vec{r}-\omega t)}.
\end{align}
The time-averaged Poynting vector of the transmitted wave reads
\begin{equation}
\vec{S}_{t}=\frac{1}{2}\mathrm{Re}(\vec{E}_{t}\times\vec{H}_{t}^{*}),
\end{equation}
with the components being
\begin{align}
S_{tx} & =\frac{1}{2}\mathrm{Re}(E_{ty}H_{tz}^{*}-E_{tz}H_{ty}^{*})\nonumber \\
 & =-\frac{1}{2}\mathrm{Re}(E_{tz}H_{ty}^{*})\nonumber \\
 & =-\frac{1}{2}E_{tz}\left(-\frac{\omega\epsilon_{z}E_{tz}}{k_{tx}}\right)\nonumber \\
 & =\frac{\omega\epsilon_{z}}{2k_{tx}}E_{tz}^{2},\\
S_{tz} & =\frac{1}{2}\mathrm{Re}(E_{tx}H_{ty}^{*}-E_{ty}H_{tx}^{*})\nonumber \\
 & =\frac{1}{2}\mathrm{Re}(E_{tx}H_{ty}^{*})\nonumber \\
 & =\frac{1}{2}E_{tx}(-\frac{\omega\epsilon_{z}E_{tz}}{k_{tx}})\nonumber \\
 & =\frac{1}{2}E_{tx}\left(-\frac{\omega\epsilon_{z}}{k_{tx}}\right)\left(-\frac{\epsilon_{x}k_{tx}E_{tx}}{\epsilon_{y}k_{tz}}\right)\nonumber \\
 & =\frac{\epsilon_{x}\omega E_{tx}^{2}}{2k_{tz}}<0,
\end{align}
because $\epsilon_{x}<0$, and $\omega,k_{tz}>0$. In order to transmit
energy from the interface into the medium, $S_{tx}$ should also be
negative, and thus $k_{tx}<0$ as $\omega,\epsilon_{z}>0$. Together
with Eq.~(\ref{eq:ExtraOrDispersion}), we have
\begin{align}
k_{tx} & =\sqrt{\mu_{0}\omega^{2}\epsilon_{z}-\frac{\epsilon_{z}}{\epsilon_{x}}k_{tz}^{2}}\nonumber \\
 & =\sqrt{\mu_{0}\omega^{2}\epsilon_{z}-\frac{\epsilon_{z}}{\epsilon_{x}}k_{iz}^{2}}\nonumber \\
 & =\sqrt{\mu_{0}\omega^{2}\epsilon_{z}-\frac{\epsilon_{z}}{\epsilon_{x}}k_{i}^{2}\sin^{2}\theta}\nonumber \\
 & =\sqrt{\mu_{0}\omega^{2}\epsilon_{z}\left(1-\frac{\epsilon_{0}}{\epsilon_{x}}\sin^{2}\theta\right)},
\end{align}
where
\begin{equation}
k_{i}^{2}=\mu_{0}\omega^{2}\epsilon_{0}.
\end{equation}

\section{Analysis of Experimental Feasibility}
\label{sec:Exp}

According to Refs.~\cite{Jackson1999,Landau1995}, the constitutive
relation reads
\begin{eqnarray}
\vec{D} & = & \epsilon_{D}\epsilon_{0}\vec{E}+\vec{P},\\
\vec{B} & = & \mu_{0}\vec{H}+\mu_{0}\vec{M},
\end{eqnarray}
where $\epsilon_{D}$ and $\mu_{0}$ are, respectively, the permittivity
and permeability of diamond without NV centers, $\vec{E}$ and $\vec{B}$
are respectively the electric and magnetic fields with frequency $\omega$,
the electric polarization and magnetization are respectively
\begin{eqnarray}
\vec{P} & = & -\frac{n_{0}}{\hbar }\mathrm{Re}\sum_{j,i,f}\rho_{i}\frac{\vec{d}_{if}(\vec{d}_{fi}\cdot\vec{E})}{\omega-\Delta_{fi}+i\gamma},\\
\vec{M} & = & -\frac{\mu_{0}n_{0}}{\hbar }\mathrm{Re}\sum_{j,i,f}\rho_{i}\frac{\vec{m}_{if}(\vec{m}_{fi}\cdot\vec{H})}{\omega-\Delta_{fi}+i\gamma}.
\end{eqnarray}
Here $\vec{d}_{if}=\langle i\vert\vec{d}\vert f\rangle$ is the matrix
element of electric dipole between the initial state $\vert i\rangle$
and the final state $\vert f\rangle$; $\sum_{j}$ is the summation
over all NV centers within the volume $v_{0}$. $\Delta_{fi}$ is
the transition frequency between the initial and final states; $\gamma$
is the homogeneous lifetime of all excited states; In the summation
the initial and final states should be different, $i\neq f$. And
the system is initially in a state with density matrix $\rho(0)=\sum_{i}\rho_{i}\vert i\rangle\langle i\vert$.

When there is no magnetic field, the Hamiltonians of the electronic
ground and excited states are respectively simplified as
\begin{eqnarray}
H_{\mathrm{gs}} & = & D_{\mathrm{gs}}S_{z}^{2}= D_{\mathrm{gs}}\sum_{m_{z}=\pm1}\vert\Phi_{A_{2};1,m_{s}}^{c}\rangle\langle\Phi_{A_{2};1,m_{s}}^{c}\vert\\
H_{\mathrm{es}} & = & D_{\mathrm{es}}^{\parallel}S_{z}^{2}+\xi(S_{y}^{2}-S_{x}^{2})
  \simeq  \sum_{\alpha}D_{\mathrm{es}}^{\parallel}\sum_{m_{z}=\pm1}\vert\Phi_{E,\alpha;1,m_{s}}^{c}\rangle\langle\Phi_{E,\alpha;1,m_{s}}^{c}\vert,
\end{eqnarray}
where we have dropped the interaction term in order to roughly estimate
the minimum density of NV centers in order to realize negative refraction.

The selection rule of optical transition is summarized as \cite{Acosta2011}
$\vert\Phi_{A_{2};S,m_{s}}^{c}\rangle\underset{\rightleftharpoons}{\alpha^{\prime}}\vert\Phi_{E,\alpha;S,m_{s}}^{c}\rangle$,
where $m_{s}=0,\pm1$, both $S$ and $m_{s}$ are conserved. The transition
electric dipole has been experimentally estimated as $11$ D \cite{Lenef1996}.
In order to qualitatively estimate the minimum density of NV centers
for realizing negative refraction, without loss of generality, the
orientations of all NV centers are assumed to be along the $z$-axis.
Thus, all of the matrix elements of the transition electric dipole
are equal
\begin{equation}
\langle\Phi_{E,\alpha;1,m_{s}}^{c}\vert\vec{d}\vert\Phi_{A_{2};1,m_{s}}^{c}\rangle=11(\hat{e}_{x}+\hat{e}_{y})\mathrm{D}.
\end{equation}
For a specific NV center, $\hat{e}_{z}$ should be replaced by the
orientation of its principle axis in the lab coordinate system, i.e.,
$\vec{r}_\mathrm{AO}$, $\vec{r}_\mathrm{BO}$, $\vec{r}_\mathrm{CO}$, and $\vec{r}_\mathrm{DO}$. Initially,
the NV center is in the state
\begin{equation}
\rho(0)=\frac{1}{3}\sum_{m_{s}=0,\pm1}\vert\Phi_{A_{2};1,m_{s}}^{c}\rangle\langle\Phi_{A_{2};1,m_{s}}^{c}\vert.
\end{equation}
Therefore,
\begin{eqnarray}
\sum_{i,f}\vec{d}_{if}\vec{d}_{fi} & = & \frac{4}{3}\times121\times(\hat{e}_{x}\hat{e}_{x}+\hat{e}_{y}\hat{e}_{y}+\hat{e}_{x}\hat{e}_{y}+\hat{e}_{y}\hat{e}_{x})\mathrm{D}^{2},\\
\vec{P} & = & -\frac{n_{0}}{\hbar}\mathrm{Re}\sum_{j,i,f}\rho_{i}\frac{\vec{d}_{if}(\vec{d}_{fi}\cdot\vec{E})}{\omega-\Delta_{fi}+i\gamma}\nonumber \\
 & = & -2\epsilon_0\zeta\gamma\mathrm{Re}\frac{(\hat{e}_{x}\hat{e}_{x}+\hat{e}_{y}\hat{e}_{y}+\hat{e}_{x}\hat{e}_{y}+\hat{e}_{y}\hat{e}_{x})\vec{E}}{\omega-\Delta_{fi}+i\gamma},\\
\zeta & = & \frac{242n_{0}\;\mathrm{D}^{2}}{9\hbar\gamma\epsilon_0},
\end{eqnarray}
where $\gamma^{-1}=10$~ns \cite{Acosta2011}, $\epsilon_{D}=5.7$
\cite{Fontanella1977} and $\mu_{D}=1-2.1\times10^{-5}$ \cite{Young1992}
are the relative permittivity and permeability of the pure diamond.
For the diamond with NV centers, the electric displacement is
\begin{eqnarray}
\vec{D} & = & \epsilon_{D}\epsilon_{0}\vec{E}+\vec{P}\nonumber \\
 & = & \left[\epsilon_{D}-2\zeta\gamma\mathrm{Re}\frac{(\hat{e}_{x}\hat{e}_{x}+\hat{e}_{y}\hat{e}_{y}+\hat{e}_{x}\hat{e}_{y}+\hat{e}_{y}\hat{e}_{x})}{\omega-\Delta_{fi}+i\gamma}\right]\epsilon_{0}\vec{E}\nonumber \\
 & = & \left[\epsilon_{D}-2\zeta\gamma\frac{(\omega-\Delta_{fi})(\hat{e}_{x}\hat{e}_{x}+\hat{e}_{y}\hat{e}_{y}+\hat{e}_{x}\hat{e}_{y}+\hat{e}_{y}\hat{e}_{x})}{(\omega-\Delta_{fi})^{2}+\gamma^{2}}\right]\epsilon_{0}\vec{E}\nonumber \\
 & \geq & \left[\epsilon_{D}-2\zeta\gamma\frac{(\hat{e}_{x}\hat{e}_{x}+\hat{e}_{y}\hat{e}_{y}+\hat{e}_{x}\hat{e}_{y}+\hat{e}_{y}\hat{e}_{x})}{2\gamma}\right]\epsilon_{0}\vec{E}
\end{eqnarray}
and thus the tensor of relative permittivity is
\begin{equation}
\overleftrightarrow{\epsilon_{r}}=\begin{pmatrix}\epsilon_{D}-\zeta & -\zeta & 0\\
-\zeta & \epsilon_{D}-\zeta & 0\\
0 & 0 & \epsilon_{D}
\end{pmatrix}
\end{equation}
with three principal components being $\epsilon_{r}^{(1)}=\epsilon_{D}-2\zeta$,
and $\epsilon_{r}^{(2)}=\epsilon_{r}^{(3)}=\epsilon_{D}$. In order to make $\epsilon_{r}^{(1)}=0$, the critical density of the NV centers is
\begin{equation}
n_{0}^{c}=\frac{9\hbar\gamma\epsilon_{D}}{\epsilon_{0}\times484\;\mathrm{D}^{2}}=1.77\times10^{21} \;\textrm{m}^{-3}.
\end{equation}
Because as shown in Fig.~\ref{fig:Orientation} two carbon atoms occupy the volume
\begin{equation}
v=\left(\frac{2}{\sqrt{3}}d\right)^{3}=(1.78\times10^{-10})^{3}\;\mathrm{m}^{3},
\end{equation}
the minimum density of the NV centers to achieve negative refraction is
\begin{equation}
\frac{1}{2}vn_{0}^{c}=5.00\;\mathrm{ppb},
\end{equation}
which is within the range of experimental fabrication, e.g. 16~ppm
\cite{Jarmola2012}.

For the NV center with the symmetry axis along $\vec{r}_\textrm{OA}$,
the principal axis of the negative permittivity is along $R(\vec{n}_\textrm{OA},\theta_\textrm{OA})(\vec{e}_x+\vec{e}_y)/\sqrt{2}$.
For the NV center with the symmetry axis along $\vec{r}_\textrm{OB}$,
the principal axis of the negative permittivity is along $R(\vec{n}_\textrm{OA},-\theta_\textrm{OA})(\vec{e}_x+\vec{e}_y)/\sqrt{2}$.
For a medium with the above two orientations,
the principal axis of the negative permittivity is along the $z$-axis.
In the same way, we can prove that for a medium with the other two orientations,
i.e. the symmetry axis along $\vec{r}_\textrm{OC}$ and $\vec{r}_\textrm{OD}$,
the principal axis of the negative permittivity is also along the $z$-axis.
Therefore, for the diamond with NV centers along the four possible orientations,
the principal axis of the negative permittivity is along the $z$-axis.